\documentclass[lettersize,journal]{IEEEtran}
\usepackage{amsmath,amsfonts}
\usepackage{algorithmic}
\usepackage{algorithm}
\usepackage{array}
\usepackage[caption=false,font=normalsize,labelfont=sf,textfont=sf]{subfig}
\usepackage{textcomp}
\usepackage{stfloats}
\usepackage{url}
\usepackage{verbatim}
\usepackage{graphicx}
\usepackage{cite}
\usepackage{svg}
\usepackage{cuted}
\usepackage{amssymb}
\DeclareMathOperator*{\argmin}{arg\,min}
\usepackage{tikz}
\usetikzlibrary{positioning}
\usetikzlibrary{calc} 
\usetikzlibrary{shapes.geometric} 
\usepackage{orcidlink}

\usepackage{hhline}
\usepackage{multirow}

\usepackage{amsthm}  

\begin{document}

\title{Dual-Polarized Rate-Splitting Multiple Access for mmWave Communications: A Practical Design and Prototype Implementation}

\author{Álvaro Pendás-Recondo\orcidlink{0000-0002-8029-8024}, Álvaro F. Vaquero\orcidlink{0000-0001-7317-4122}, Jesús Alberto López-Fernández\orcidlink{0000-0001-7603-9591} and Marcos R. Pino\orcidlink{0000-0001-7468-3084}
\thanks{This work was supported in part by Grant PID2023-146246OB-C32 (ANT4IT) funded by Spanish Government through MCIU/AEI/10.13039/501100011033/FEDER, UE; and in part by Government of Asturias-Sekuens/FEDER under grants IDE-2024-000693, and “Severo Ochoa” Program PA-22-BP21-116.

Álvaro Pendás-Recondo, Álvaro F. Vaquero, Jesús Alberto López-Fernández, and Marcos R. Pino are with the Department of Electrical Engineering, Universidad de Oviedo, 33203, Gijón, Spain (e-mail: \{pendasalvaro,  fernandezvalvaro; jelofer; mpino\}@uniovi.es). They all belong to the Signal Theory and Communications Group.}}

\markboth{}%
{Shell \MakeLowercase{\textit{et al.}}: A Sample Article Using IEEEtran.cls for IEEE Journals}


\maketitle

\begin{abstract}

This work presents a dual-polarized rate-splitting multiple access (RSMA) design for millimeter-wave (mmWave) or frequency range 2 (FR2) communications. Focusing on a practical implementation, we consider the specific scenario of a single transmitter serving two receivers (users) within the same time, frequency, and code resource block. The transmitter and each receiver are equipped with a pair of dual-polarized antennas. We first present the system formulation and compute the achievable rate expressions for discrete constellations. Based on these expressions, we propose and evaluate two receiver architectures: one based on successive interference cancellation (SIC) and a non-SIC decoder. Both designs exhibit comparable performance, and the combination of polarization diversity and RSMA provides the system with significant flexibility to adapt to different channel conditions when serving both users, while the non-SIC decoder enables a simpler hardware implementation. We then present a prototype implementation of the proposed design, operating at a carrier frequency of $25$ GHz, and evaluate it by measuring the symbol error rate (SER) of the different data streams under various signal-to-noise ratio (SNR) and polarization diversity conditions. The obtained measurements show good agreement with software simulations, thereby validating the system implementation. To the best of our knowledge, this is the first mmWave RSMA prototype reported in the literature and the first to exploit polarization diversity. Furthermore, given the selected carrier frequency, the proposed design may also find applicability in the upper portion of the envisioned FR3 band (7–24 GHz).

\end{abstract}


\section{Introduction}
\IEEEPARstart{T}he study of new multiple access (MA) technologies for wireless communications, largely based on concepts of information theory known for decades, has garnered considerable attention in recent years. Space-division multiple access (SDMA), which exploits multiple-input multiple-output (MIMO) or multiple-input single-output (MISO) architectures for MA, has already been successfully implemented and deployed in many wireless standards, providing a considerable leap in spectral efficiency compared with single-input single-output (SISO). Beyond SDMA, power-domain non-orthogonal multiple access (NOMA), which is capacity-achieving in the SISO broadcast channel, has demonstrated strong potential for exploiting channel strength diversity among users \cite{ojcom}. However, NOMA has been shown to scale less efficiently than SDMA with an increasing number of antennas \cite{noma_miso}. Moreover, practical deployment challenges, including the use of successive interference cancellation (SIC) receivers \cite{noma_imp1}, have led to NOMA being considered in the 5G standard but not widely deployed to date. More recently, rate-splitting multiple access (RSMA) \cite{rsma_survey1}, \cite{rsma_survey2} has been proposed as a new framework for managing inter-user interference by combining principles from SDMA, power-domain NOMA, and the rate-splitting concept originally proposed in \cite{HK}, offering a flexible yet powerful MA strategy.  RSMA includes SDMA and power-domain NOMA as particular cases and, more importantly, has been shown to outperform both in many scenarios \cite{rs_letter} while providing enhanced robustness to imperfect channel state information at the transmitter (CSIT) \cite{imperfect_csi1}, \cite{imperfect_csi2}, \cite{imperfect_csit3}. Overall, these considerations make RSMA especially appealing as a physical-layer MA strategy for future wireless standards and technologies, particularly, although not exclusively, for the downlink \cite{rsma_d1}, \cite{rsma_d2}. More recently, the combination of RSMA with polarization diversity has also been explored theoretically \cite{dualRSMA}, \cite{dualRSMA2}, \cite{dualRSMA3}, \cite{dualRSMA4}.

Within the same context of seeking new solutions to address the exponential growth in wireless devices, the adoption of millimeter-wave (mmWave) bands, also known as frequency range 2 (FR2) in 5G (24.25 - 71 GHz), provides access to a largely underutilized portion of the spectrum, enabling higher data rates and lower latency through larger bandwidths than those available in sub-6 GHz bands. Nonetheless, the well-known challenges associated with these bands, including severe propagation and penetration losses, as well as increased cost and complexity of hardware architectures, channel estimation, and synchronization \cite{mmWave1}, \cite{mmWave2}, \cite{mmWave3}, \cite{mmWave4} have slowed their widespread adoption. These limitations have even motivated discussions on the use of the upper-mid (or FR3) band for 6G and future wireless standards, roughly from 7 GHz to 24 GHz, as a trade-off between implementation complexity and spectrum resources.

\subsection{RSMA Prototype Implementations}
Table \ref{table_rsma} presents previous RSMA prototype implementations, all of which are based on software-defined radio (SDR), indicating the corresponding reference, year of publication, center frequency, number of transmit antennas, and number of receive antennas per user. We note that practical RSMA implementations are relatively recent, with the first reported in \cite{rsma_imp1}. The work in \cite{rsma_imp2} investigates RSMA in SISO configurations; \cite{rsma_imp3} focuses on a multi-group multicast scenario; and \cite{rsma_imp4} addresses the practical application of RSMA to integrated sensing and communications (ISAC). Finally, \cite{rsma_mag} reviews these prototype efforts, presents further results, and discusses the potential of RSMA as a natural successor to SDMA in 6G. To the best of our knowledge, our work presents the first implementation of RSMA in the mmWave band, the first to employ multiple receive antennas per user, and the first to combine RSMA with polarization diversity, an idea that has been previously proposed theoretically \cite{dualRSMA} for Gaussian inputs, or Gaussian signaling (GS).

\begin{table}[t]
\caption{RSMA prototype implementations.}
\centering
\begin{tabular}{|c|c|c|c|c|}
\hline
Reference & Year & Frequency & Tx. Antennas & Rx. Antennas \\ \hline \hline

\cite{rsma_imp1} & 2024 & $2.484$ GHz & $2$ & $1$  \\ \hline

\cite{rsma_imp2} & 2024 & $2.4$ GHz & $1$ & $1$  \\ \hline

\cite{rsma_imp3} & 2025 & $2.484$ GHz & $2$ & $1$  \\ \hline

\cite{rsma_imp4} & 2025 & $2.484$ GHz & $2$ & $1$  \\ \hline

\cite{rsma_mag} & 2025 & $2.5$ GHz & $2$ & $1$  \\ \hline

\textbf{This work} & 2026 & $25$ GHz & $2$ & $2$  \\ \hline

\end{tabular}
\label{table_rsma}
\end{table}

\subsection{Motivation: mmWave and Dual-Polarized RSMA}
The challenges associated with mmWave bands, including accurate channel state information (CSI) acquisition, hardware impairments, and system synchronization, motivate the exploration of solutions that can exploit the abundant spectral resources of these bands as well as the  diversity offered by MIMO, while maintaining a simple hardware architecture. Motivated by these considerations, we present the practical combination of polarization diversity with RSMA to serve multiple users with enhanced spectral efficiency without relying on instantaneous CSIT, which is particularly difficult to obtain in FR2 bands \cite{csitm}.

We consider the specific scenario in which the transmitter and each receiver are equipped with a pair of dual-polarized antennas, with two users being served within the same time, frequency, and code resource block. Since the receivers’ positions and orientations may vary over time, their channel strengths and degrees of polarization diversity may also vary. However, these parameters are expected to change more slowly than the instantaneous channel phases, especially in the mmWave band, due to the small wavelength. Therefore, the transmitter is assumed to have knowledge only of the channel strength and polarization diversity of each receiver, which can be viewed as a particular instance of magnitude CSIT \cite{vigo}. Furthermore, the proposed design requires no phase synchronization among the transmit antennas, although they must be synchronized in time to enable collaboration.

According to RSMA, the messages of each user are split into common and private parts. The common messages are combined into a single common stream, resulting in a total of three streams (one common and two private) to be transmitted. The proposed design consists of using one polarization to transmit the common stream and the other polarization to transmit a superposition of the private streams. This design is inspired by one of the proposals presented in \cite{dualRSMA} for GS. However, since the objective is to present a practical implementation, we model the system using finite input constellations, also known as discrete signaling (DS) \cite{ojcom}. This modeling significantly affects the choice of superposition coding techniques at the transmitter (combined with constellation selection) \cite{ojcom}, \cite{Mod_Cod_NOMA}, as well as the design of receiver architectures, including SIC, joint maximum-likelihood decoding (JMLD) or joint de-mapping \cite{rsma_finite}, and treating discrete interference as noise (TIN) \cite{tin1}. Apart from terrestrial communications, the proposed design might also find application in satellite communications (SatCom) due to the considered frequency bands, channel conditions, and the limited number of transmit antennas in the downlink.

\subsection{Contributions}
The main contributions of this work are summarized as follows:

\begin{itemize}
\item We present an integration of polarization diversity and RSMA in the mmWave band for the specific scenario of a single transmitter and two receivers, all equipped with a pair of dual-polarized antennas. Although some related concepts were previously proposed in \cite{dualRSMA}, we base our model on DS rather than GS, focusing on the practical system implementation. This choice has important implications for system design, particularly regarding superposition coding at the transmitter and decoding strategies at the receiver. Furthermore, motivated by the characteristics of mmWave channels, the transmitter operates using only knowledge of the channel strength and polarization diversity, without requiring instantaneous phase CSIT.

\item We propose two different receiver architectures: one that combines SIC with JMLD; and a non-SIC receiver based only on JMLD and TIN decoding. To compare both strategies, we derive mutual information (or constellation-constrained capacity) expressions to evaluate the achievable rates. Based on these expressions and Monte Carlo simulations, we provide a numerical comparison of both approaches across different scenarios, including different degrees of polarization diversity, disparities in channel strength among users, signal-to-noise ratio (SNR) levels, and constellation and power allocation choices. The results show that the SIC-based architecture provides only a modest increase in achievable rate, at the cost of increased hardware complexity and more demanding channel estimation requirements at the receivers. Additionally, we comment on the robustness of the proposed designs against imperfect CSIT, their scalability to accommodate more users, and their combination with other MA strategies within the flexible framework provided by RSMA.

\item Based on the previous results, we select the decoder based on the combination of JMLD and TIN, and present a practical implementation of the system operating at 25 GHz, based on SDR combined with commercial mixers, power amplifiers, and antennas. To the best of our knowledge, this work presents the first implementation of RSMA in the mmWave band and the first to combine RSMA with polarization diversity. The chosen frequency not only validates an FR2 band implementation, but also indicates that the proposed solution can be readily extended to operation in the upper portion of the FR3 band, for example, between $20$ and $24$ GHz.

\item We present experimental results of the proposed system based on the evaluation of the symbol error rate (SER) under different SNR and polarization conditions. To ensure accuracy and reproducibility, the measurement campaign is conducted in the anechoic chamber at the University of Oviedo, where the dual-polarized antennas can be positioned and rotated automatically with high precision (on the order of hundredths of a degree). The results show good agreement with Monte Carlo simulations, thereby validating the practical implementation of the system.

\end{itemize}

The remainder of this article is organized as follows. Section \ref{SM} introduces the system model for the scenario considered. Section \ref{DRSMA} proposes a specific application of dual-polarized RSMA to the problem, considering both SIC and non-SIC receivers. Section \ref{NR} presents numerical results for the achievable rate analysis, based on the mutual information computations in Appendix \ref{appa}. The practical implementation of the system is described in Section \ref{PI}, and experimental results are reported in Section \ref{ER}. Finally, conclusions are drawn in Section \ref{Con}.

\section{System Model}\label{SM}

\subsection{System Model}
Consider a transmitter communicating with two receivers within the same time, frequency and code resource block. The transmitter and each receiver are equipped with a pair of dual-polarized antennas, such that each pair is composed of two orthogonal linearly polarized antennas. Let us denote $\bold{x}\in \mathbb{C}^{2 \times 1}$ as the transmitted symbol vector, $\bold{x} = \left[ x_\textrm{V}, x_\textrm{H}\right]^T$. For simplicity, we omit the time index and assume that the transmit and receive antennas lie in the same horizontal plane; therefore, we denote the two orthogonal polarizations as horizontal and vertical. However, this modeling can be readily extended to arbitrary antenna positions. At the $i$-th receiver, $i\in\{1,2\}$, the received signal, $\bold{y}_i$, is expressed as

\begin{equation}\label{yi}
\bold{y}_i = \begin{bmatrix}
y_{i\textrm{V}} \\
y_{i\textrm{H}} 
\end{bmatrix}= \bold{H}_i\bold{x} + \bold{n}_i  = \bold{H}_i \begin{bmatrix}
x_\textrm{V} \\
x_\textrm{H} 
\end{bmatrix} + \begin{bmatrix}
n_{i\textrm{V}} \\
n_{i\textrm{H}} 
\end{bmatrix},
\end{equation} where the additive white Gaussian noise (AWGN) components per polarization, denoted as $n_{i\textrm{V}} $ and $n_{i\textrm{H}}$, are independent and identically distributed, (i.i.d.), following a zero-mean complex normal distribution, i.e., $n_{i\textrm{V}}\sim \mathcal{C}\mathcal{N}\left(0,\sigma_i^{2}\right)$, and $n_{i\textrm{H}}\sim \mathcal{C}\mathcal{N}\left(0,\sigma_i^{2}\right)$. The channel matrix, $\bold{H}_i$, is defined as 

\begin{equation}\label{Hi}
\bold{H}_i = \begin{bmatrix}
h_{i\textrm{\textrm{VV}}} & h_{i\textrm{VH}}\\
h_{i\textrm{HV}} & h_{i\textrm{HH}}
\end{bmatrix} = \begin{bmatrix}
r_{i\textrm{\textrm{VV}}} \chi_{i\textrm{\textrm{VV}}} & r_{i\textrm{VH}} \chi_{i\textrm{VH}}\\
r_{i\textrm{HV}} \chi_{i\textrm{HV}} & r_{i\textrm{HH}} \chi_{i\textrm{HH}}
\end{bmatrix},
\end{equation} where each component is obtained as the product of two elements. For instance, $h_{i\textrm{VV}} \in \mathbb{C}$ denotes the channel coefficient between the vertically polarized transmit and receive antennas for the $i$-th receiver and it is given by $h_{i\textrm{VV}} = r_{i\textrm{VV}} \chi_{i\textrm{VV}}$, where $r_{i\textrm{VV}} \in \mathbb{C}$ represents the channel scaling and rotation of the transmitted signal, and $\chi_{i\textrm{VV}} \in \mathbb{R}$ denotes the co-polarization coefficient. Note that $\chi_{i\textrm{VV}}$ and $\chi_{i\textrm{HH}}$ are the co-polarization coefficients, while $\chi_{i\textrm{HV}}$ and $\chi_{i\textrm{VH}}$ represent the channel cross-polarization coefficients. They can be expressed as real numbers, since the phase information is captured by the corresponding $r_i$ components. The polarization coefficients lie within the interval $[0,1]$ and satisfy the normalization conditions, $\chi_{i\textrm{VV}}^2 + \chi_{i\textrm{VH}}^2 = 1$ and $\chi_{i\textrm{HH}}^2 + \chi_{i\textrm{HV}}^2 = 1$. In the case of perfectly aligned antennas, $\chi_{i\textrm{VV}}^2 = \chi_{i\textrm{HH}}^2 = 1$, leading to $h_{i\textrm{HV}} = h_{i\textrm{VH}} = 0$. We define the signal-to-cross-polarization interference discrimination (SXPD) for the $i$-th user as  $\textrm{SXPD}_{i\textrm{V}} = 20\log_{10}\left(\frac{\chi_{i\textrm{VV}}}{\chi_{i\textrm{VH}}}\right)$ and $\textrm{SXPD}_{i\textrm{H}} = 20\log_{10}\left(\frac{\chi_{i\textrm{HH}}}{\chi_{i\textrm{HV}}}\right)$, for the vertical and horizontal polarizations, respectively. For the sake of simplicity, from this point onward, a generic channel coefficient is denoted as $h_{i\textrm{PL}}$, where $\textrm{P, L} \in \{\textrm{V, H}\}$. This notation is similarly extended to the components $r_{i\textrm{PL}}$ and $\chi_{i\textrm{PL}}$. Finally, note that the proposed formulation can be readily extended to the case of two orthogonal circularly polarized antennas. We also note that other channel decompositions are possible while preserving the same channel model. The main reason for adopting the present decomposition is that it enables the effects of multiuser channel strength and polarization conditions to be analyzed independently, as shown in Section \ref{NR}.

\subsection{Channel Model}\label{CM}
We assume a flat-fading channel model with rapidly varying phase and block-static amplitude; that is, the magnitude $\left|r_{i\textrm{PL}}\right|$ and the polarization coefficients $\chi_{i\textrm{PL}}$ in Eq. (\ref{Hi}) are assumed to remain constant over a transmission block for a fixed number of symbols (e.g., 512, 1024, 4096, etc.). However, the phase component $\angle r_{i\textrm{PL}} = \phi_{i\textrm{PL}}$ varies rapidly, potentially taking a different realization for each symbol. For mutual information calculations in Section \ref{NR}, we adopt a worst-case assumption in which the phases follow i.i.d. uniform distributions for every symbol, $\phi_{i\textrm{PL}} \sim \mathcal{U}[0, 2\pi)$. However, it must be noted that even in rapidly varying environments, the channel phase typically exhibits temporal correlation, i.e., the phase depends on previous realizations \cite{timechannel}. This characteristic is exploited by the receiver in the practical implementation to track the phase for channel estimation, which in turn enables reliable decoding (see Section~\ref{PI}). Nonetheless, due to the rapidly varying phase, the transmitter is assumed to have access only to $\left|r_{i\textrm{PL}}\right|$ and $\chi_{i\textrm{PL}}$, allowing it to determine $\left|h_{i\textrm{PL}}\right|$. This form of channel knowledge falls within the category of magnitude CSIT \cite{ojcom}, \cite{vigo}.

Note that the considered channel model implicitly assumes the presence of a dominant path component and negligible multipath, while accounting for a time-varying scenario in which small positional changes can cause significant phase variations, particularly in the mmWave band. For instance, at $25$ GHz, the wavelength is $\lambda = 12$ mm, so moving the receiver by just $1$ mm results in a phase shift of up to $30^\circ$. However, the same positional change would have a negligible impact on both the channel propagation losses and the polarization coefficients in the presence of a dominant path. In case of a more dispersive channel (but still dominated by line-of-sigh (LoS) conditions) integration with OFDM, which is pointed out as future line of research in Section \ref{Con}, can transform the effective channel of each subcarrier into a single complex coefficient where the magnitude would also vary slower than the phase \cite{csitm}. Although each application and frequency band requires a detailed analysis for its specific design, these considerations generally apply to many SatCom communication scenarios \cite{satcom_sur}, as well as to the upper portion of the envisioned FR3 band (7-24 GHz), more specifically the 20-24 GHz range, where most communications would rely on LoS conditions.

Finally, it is worth mentioning that, to ensure reliable connectivity in mmWave bands despite significant penetration and propagation losses \cite{mmWave1}, \cite{mmWave2}, relaying technologies such as intelligent reflecting surfaces (IRSs) are of particular interest in both terrestrial and SatCom (downlink) scenarios, especially in urban environments, where they can help establish the aforementioned LoS conditions. These can be either fully passive surfaces with a fixed configuration to extend coverage to a specific area, also known as smart electromagnetic skins (SES) \cite{ses_ojap}, \cite{ses}, \cite{ses_mag}, or reconfigurable solutions, also known as reconfigurable intelligent surfaces (RIS) with dynamic, user-tracking capabilities \cite{ris_mm_imp2}. Although beyond the scope of this work, previous studies have proposed effective solutions for real-time RIS reconfiguration without requiring instantaneous phase CSIT, instead relying on a codebook of predefined beam patterns combined with a selection algorithm \cite{mmWave2}, \cite{bemcodebook}, \cite{codebook_tutorial}.

\section{Dual-Polarized Rate-Splitting Multiple Access}\label{DRSMA}

In this section, we present the application of single-layer RSMA to the scenario described in Section \ref{SM}. We begin by describing the transmitter, followed by two different types of receivers. We then define the performance metrics considered, namely achievable rate, calculated as the mutual information between the transmitted and received streams, and SER.

\subsection{Rate-Splitting Multiple Access, Transmitter} \label{txdes}
In the ideal scenario of perfectly polarized and aligned antennas, with $\chi_{i\textrm{VV}} = \chi_{i\textrm{HH}} = 1$, the system would achieve two completely independent streams (two degrees of freedom). In the considered two-user scenario, it would be straightforward to assign one polarization per user, thereby completely eliminating inter-user interference and requiring only one antenna at each receiver. However, this assumption is unrealistic in the presence of mobile users, whose position and orientation are subject to change over time. To select an MA strategy capable of serving both users effectively while exploiting the polarization state at any moment for maximum spectral efficiency, we propose the use of single-layer RSMA \cite{rsma_survey1}, \cite{rsma_survey2} combined with polarization diversity.

Considering single-layer RSMA, at the transmitter, the data to be sent to each user is split into common and private parts. The common parts from both users are combined into a common message, resulting in a total of three data streams (one common and two private). The common stream, intended to be decoded by both users, is denoted by $s_0$, while the private stream for the $i$-th user is denoted by $s_i$. Thus,  $s_k\in \mathbb{C}$ refers to the $k$-th stream, with $k=\{0,1,2\}$, and the $i$-th user, $i\in\{1,2\}$, intends to decode $s_0$ and $s_i$. Following one of the proposals presented in \cite{dualRSMA} for GS, adapted here for DS, the common stream is sent by one of the antennas, in this case, the vertically polarized, while a superposition of both private streams is sent by the horizontally polarized, which results in

\begin{equation}\label{tx1}
\begin{cases}
\begin{aligned}
&x_{\textrm{V}} = \sqrt{P_0}s_0\\
&x_{\textrm{H}} = \sqrt{P_1}s_1 + \sqrt{P_2}s_2,
\end{aligned}
\end{cases}
\end{equation} where $P_k$ denotes the power allocated to the $k$-th stream. According to an average power constraint, $P_0+P_1+P_2=P_{\textrm{t}}$, being $P_{\textrm{t}}$ the total power, while we define the coefficients $\gamma_k=\frac{P_k}{P_{\textrm{t}}}$.

Considering DS, or finite input constellations for RSMA \cite{rsma_finite}, \cite{rsma_finite2}, \cite{rsma_finite3}, \cite{rsma_finite4}, $s_k$, used to denote the $k$-th stream and its associated random variable, takes values from the alphabet $\mathcal{S}_k$, given by

\begin{equation}\label{sets1}
\mathcal{S}_k = \{ s_{k1}, s_{k2}, \ldots, s_{kM_k}\}\\,
\end{equation} where $M_k$ denotes the cardinality of the $k$-th alphabet, $M_k = \left|\mathcal{S}_k\right|$. We refer to $\mathcal{S}_k$ as the source constellation used by the $k$-th stream (or simply the constellation), while the transmitted constellation for the $k$-th stream is denoted by $\mathcal{X}_k$, with $\mathcal{X}_k = \sqrt{P_k}\mathcal{S}_k$. Accordingly, $x_k = \sqrt{P_k}s_k$, for $k \in \{0,1,2\}$.

For the superposed private streams, it is useful to define an additional alphabet that results from the mapping  $\psi: \mathcal{X}_{1} \times \mathcal{X}_{2} \to \mathcal{W}$. The resulting random variable and stream are denoted as $w$ and given by $w=x_1+x_2=\sqrt{P}_1s_1 + \sqrt{P}_2s_2$,  whose possible realizations and elements of the superposed alphabet are expressed as $w_{ml}=\psi\left(s_{1m},s_{2l}, P_1, P_2\right) = \sqrt{P_1}s_{1m} + \sqrt{P_2}s_{2l}$. Assuming that every symbol in the superposed constellation is unique, we have $\left|\mathcal{W}\right| = \left|\mathcal{X}_{1}\right| \left|\mathcal{X}_{2}\right|= M_1 M_2$. For simplicity, we define $M_{12} = M_1 M_2$. With these considerations, Eq. (\ref{tx1}) can be rewritten as

\begin{equation}\label{tx2}
\begin{cases}
\begin{aligned}
&x_{\textrm{V}} = x_0\\
&x_{\textrm{H}} = x_1+x_2=w.
\end{aligned}
\end{cases}
\end{equation}

Note that the proper selection of constellations and power allocation is crucial for obtaining a suitable superposed constellation \cite{Mod_Cod_NOMA}, and it is carried out at the transmitter based on the available magnitude and polarization CSIT, as detailed in Section \ref{CM}. 

\subsection{SIC Receiver}\label{SIC}
Two different receivers are proposed in this section: one that relies on SIC, presented here, and another that operates without SIC, introduced in Section \ref{NSIC}. Apart from SIC, both receivers implement JMLD, or joint de-mapping, and TIN.

Based on Eq. (\ref{yi}), Eq. (\ref{Hi}) and Eq. (\ref{tx2}), the received signals per polarization at the $i$-th user are given by

\begin{equation}\label{yi_sisoic}
\begin{cases}
\begin{aligned}
&y_{i\textrm{V}} = x_0h_{i\textrm{VV}} + wh_{i\textrm{VH}} + n_{i\textrm{V}}\\
&y_{i\textrm{H}} = wh_{i\textrm{HH}} + x_0h_{i\textrm{HV}} + n_{i\textrm{H}},
\end{aligned}
\end{cases}
\end{equation} where, by implementing SIC at each user, the common stream is decoded first, as it is standard in RSMA \cite{rsma_survey2}, \cite{rsma_finite}. In this context, the decoding process begins with the vertical polarization, where the common stream is decoded while treating the private streams as noise (TIN decoding). Subsequently, the horizontal polarization is decoded after subtracting the estimated common symbol, denoted by $\hat{x}_{i0}$. Accordingly, the decoding process at the $i$-th user can be expressed as

\begin{equation}\label{sic_decod}
\begin{cases}
\begin{aligned}
&\hat{x}_{i0} = \argmin_{(x_{0n})} \{\left| y_{i\textrm{V}} - x_{0n}h_{i\textrm{VV}} \right|\}\\
&\hat{w}_{i} = \argmin_{(w_{ml})} \{\left| y_{i\textrm{H}} - \hat{x}_{i0}h_{i\textrm{HV}} - w_{ml}h_{i\textrm{HH}} \right|\},
\end{aligned}
\end{cases}
\end{equation} where, using the superposed constellation as a reference for decoding the private streams, JMLD is implemented. From the estimated superposed symbol, $\hat{w}_{i}$, the first user recovers the index $m$, whereas the second user recovers the index $l$.

It is important to note that, although the decoder is described mathematically using a symbol-by-symbol approach, best practice for SIC involves decoding the $x_0$ symbols in a block. Only after applying error correction to the entire block should the receiver proceed with SIC to decode the private streams. Although this strategy reduces error propagation, it is also the primary source of the increased latency and hardware complexity associated with SIC in practical implementations \cite{noma_imp1}. On the other hand, while a hard decoder is described here, SIC based on soft decoding is also feasible. Further details and additional considerations for implementing SIC in RSMA systems can be found in \cite{rsma_finite} and \cite{Mod_Cod_NOMA}.

The SIC receiver is optimal for GS and infinite block length, assuming the achievable rate for decoding the common stream exceeds that of any private stream prior to SIC \cite{book_in}. Under these conditions, its main advantage is the complete removal of interference from the common stream to the private streams. However, under DS and finite block length, preventing error propagation requires a system design that ensures the interference from the private streams to the common stream is sufficiently reduced. This constraint limits flexibility in system design, particularly in terms of power allocation and constellation selection \cite{sic_jmld}. Apart from the increased latency resulting from the sequential nature of SIC decoding, the SIC decoder considered here has the additional drawback of requiring knowledge of $h_{i\textrm{HV}}$. While any coherent receiver must at least estimate $h_{i\textrm{VV}}$ and $h_{i\textrm{HH}}$, acquiring one or more of the cross-polarization channel coefficients further increases overall system complexity.

\subsection{Non-SIC Receiver}\label{NSIC}
The receiver without SIC can be straightforwardly obtained by modifying Eq. (\ref{sic_decod}) into

\begin{equation}\label{nosic_decod}
\begin{cases}
\begin{aligned}
&\hat{x}_{i0} = \argmin_{(x_{0n})} \{\left| y_{i\textrm{V}} - x_{0n}h_{i\textrm{VV}} \right|\}\\
&\hat{w}_{i} = \argmin_{(w_{ml})} \{\left| y_{i\textrm{H}} - w_{ml}h_{i\textrm{HH}} \right|\},
\end{aligned}
\end{cases}
\end{equation} where the decoding process now treats the common and private streams, and thus the information carried in both polarizations, independently. In both cases, TIN is applied, treating the signal coming from the cross-polarization as noise. This comes at the cost of the common stream interfering with JMLD decoding of the private streams, unlike in the SIC-based receiver. Consequently, system design regarding power allocation and constellation selection must be carefully adapted to the channel and polarization conditions, which determine the level of cross-interference. An important advantage is that this architecture requires the receiver to know only $h_{i\textrm{VV}}$ and $h_{i\textrm{HH}}$, which enhances the independence of each polarization chain and further simplifies implementation. Further details regarding TIN decoding, specifically treating discrete interference as AWGN noise, can be found in \cite{tin1} and \cite{tin2}. Although treating discrete interference as Gaussian noise is suboptimal, discrete interference has the advantage of being amplitude-bounded compared to Gaussian interference, which can, in some cases, lead to improved performance for a TIN decoder when comparing GS against DS. Finally, it should be noted that, although non-SIC receivers generally tend to have a higher computational cost \cite{rsma_finite}, the specific SIC and non-SIC decoders presented in Eq. (\ref{sic_decod}) and Eq. (\ref{nosic_decod}), respectively, have the same asymptotic computational complexity, namely, $\mathcal{O}(M_0 + M_1M_2)$.

\subsection{Achievable Rate and Symbol Error Rate}\label{FM}

We now introduce the figures of merit used to evaluate system performance: achievable rate and SER. Here, achievable rate refers to the constellation-constrained capacity \cite{rsma_finite}, i.e., the mutual information between the channel input and output with DS. This metric is used in Section \ref{NR} to guide system design. SER, on the other hand, is used in Section \ref{ER} to evaluate the practical system implementation.

The achievable rate of the $k$-th stream for the $i$-th receiver is denoted as $R_{ik}$ and defined as

\begin{equation}\label{sets2}
R_{ik}= I(y_{i\textrm{P}};s_k)\\,
\end{equation} where $I(y_{i\textrm{P}};s_k)$ denotes the mutual information between $y_{i\textrm{P}}$ and $s_k$, with $\textrm{P}\in\{\textrm{V, H}\}$ ($\textrm{P}=\textrm{V}$ when $k=0$, $\textrm{P}=\textrm{H}$ when $k=1,2$). The total achievable rate for the $i$-th user, $R_i$, is calculated as 

\begin{equation}\label{ri}
R_i = \beta_i\min\left(R_{10}, R_{20}\right) + R_{ii},\\
\end{equation} where $\beta_i$, $0\leq\beta_i\leq1$, $\beta_1+\beta_2=1$, is a factor that denotes the portion of the common stream allocated to the $i$-th user. Note that minimum value between $R_{10}$ and $R_{20}$ is chosen because the common stream must be decoded by both users \cite{rsma_survey1}, \cite{rsma_survey2}. Consequently, the total achievable rate of the system, denoted as $R_\textrm{sum}$, is obtained as 

\begin{equation}\label{rt}
R_{\textrm{sum}} = \min\left(R_{10}, R_{20}\right) + R_{11} + R_{22}.\\
\end{equation}

For notational simplicity, we denote the achievable rates computed using the non-SIC receiver simply as $R_{ik}$, $R_i$, and $R_{\textrm{sum}}$, and those computed using the SIC receiver as $R_{ik}^{\textrm{SIC}}$, $R_i^{\textrm{SIC}}$, and $R_{\textrm{sum}}^{\textrm{SIC}}$. The expressions used to evaluate these rates numerically are computed in Appendix \ref{appa}. Note that different assumptions are adopted for the two decoders based on our goal of drawing conclusions regarding the specific practical designs considered here rather than providing a general comparison between SIC and non-SIC architectures for RSMA with DS, addressed in \cite{rsma_finite}.

Regarding error performance, deriving a closed-form expression for the symbol error probability of each stream is highly challenging for arbitrary constellations. However, it can be estimated through SER measurements, either via Monte Carlo simulations or practical hardware implementation, using the decoders defined in Eq. (\ref{sic_decod}) and Eq. (\ref{nosic_decod}). The resulting SER at the $i$-th user for the $k$-th stream is denoted by $\textrm{SER}_{ik}$.

\section{Numerical Results and System Design}\label{NR}

In this section, we present insights for system design, focusing on power allocation, constellation selection, and receiver type. Additionally, we discuss the robustness of the proposed design against imperfect CSIT, compare its performance with that of alternative approaches, and comment on its scalability to support a larger number of users. These insights are derived from numerical evaluations of the rate formulas presented in Section \ref{FM} and Appendix \ref{appa}. The practical system implementation, detailed in Section \ref{PI}, is guided by the findings from this analysis. 

\subsection {Numerical Results and System Design, Similar Channels}\label{NR_SC}
We begin by analyzing a scenario in which both users experience similar channel strength. To this end, we set $\left|r_{1\textrm{PL}}\right| = \left|r_{2\textrm{PL}}\right| = 1$ in Eq. (\ref{Hi}), and consider various combinations of cross-polarization interference levels, assuming the same level applies to both antennas of the same user, $\textrm{SXPD}_{i\textrm{V}}=\textrm{SXPD}_{i\textrm{H}}=\textrm{SXPD}_{i}$, and to both users, $\textrm{SXPD}_{1}=\textrm{SXPD}_{2}$. The noise variances are set to $\sigma_1^2 = \sigma_2^2 = 1$, and the transmission SNR is defined with respect to the total power as $\textrm{SNR} = \displaystyle \frac{P_{\textrm{t}}}{\sigma_i^2} = P_{\textrm{t}}$. Regarding channel phase, for the four channel phase coefficients, $\phi_{i\textrm{PL}}$, we assume i.i.d. uniform distributions, $\phi_{i\textrm{PL}}\sim \mathcal{U}[0,2\pi)$.

\begin{figure*}
\centering
\subfloat[\normalsize $\textrm{SXPD}_1=\textrm{SXPD}_2=20 \; \textrm{dB},\; \gamma_1=\gamma_2$ \label{fr1_1}]{%
      \includegraphics[width=0.33\linewidth]{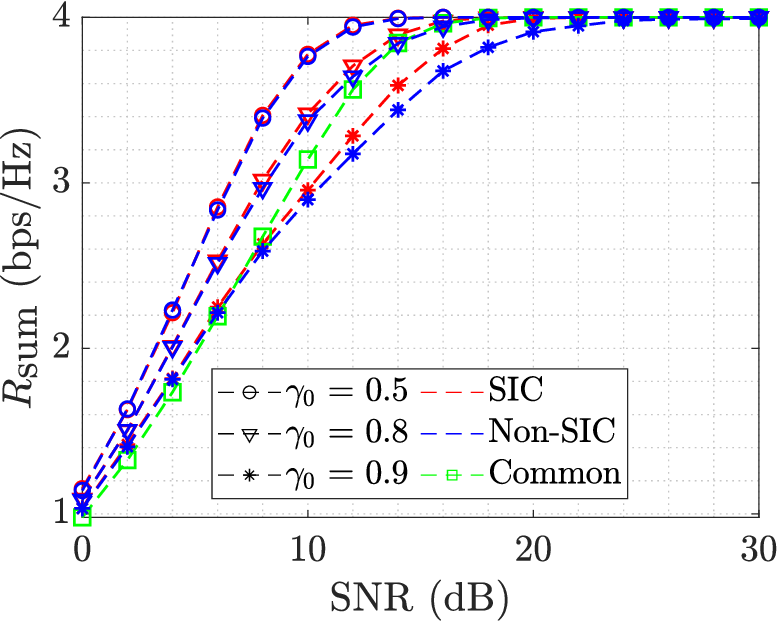}}
      \hfill
\subfloat[\normalsize  $\textrm{SXPD}_1=\textrm{SXPD}_2=10 \; \textrm{dB},\; \gamma_1=\gamma_2$ \label{fr1_2}]{%
      \includegraphics[width=0.33\linewidth]{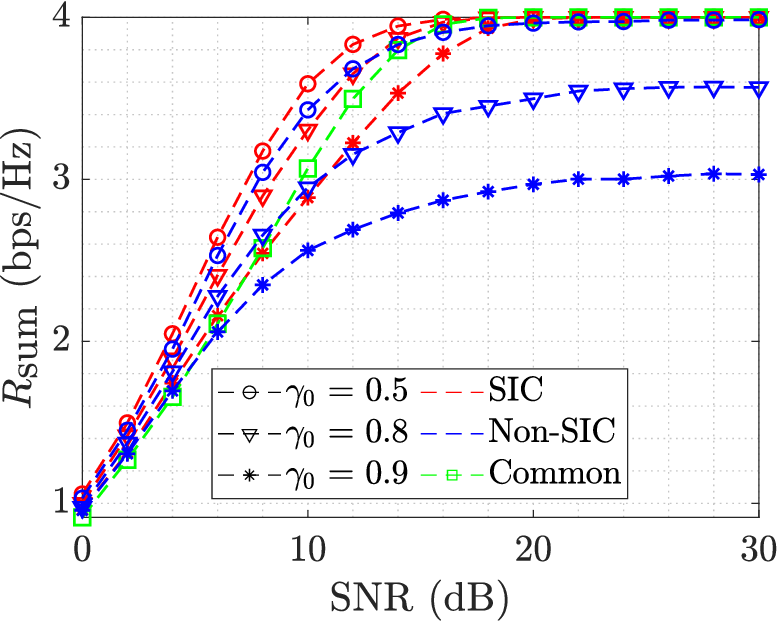}}
        \hfill
\subfloat[\normalsize  $\textrm{SXPD}_1=\textrm{SXPD}_2=3 \; \textrm{dB},\; \gamma_1=\gamma_2$ \label{fr1_3}]{%
      \includegraphics[width=0.33\linewidth]{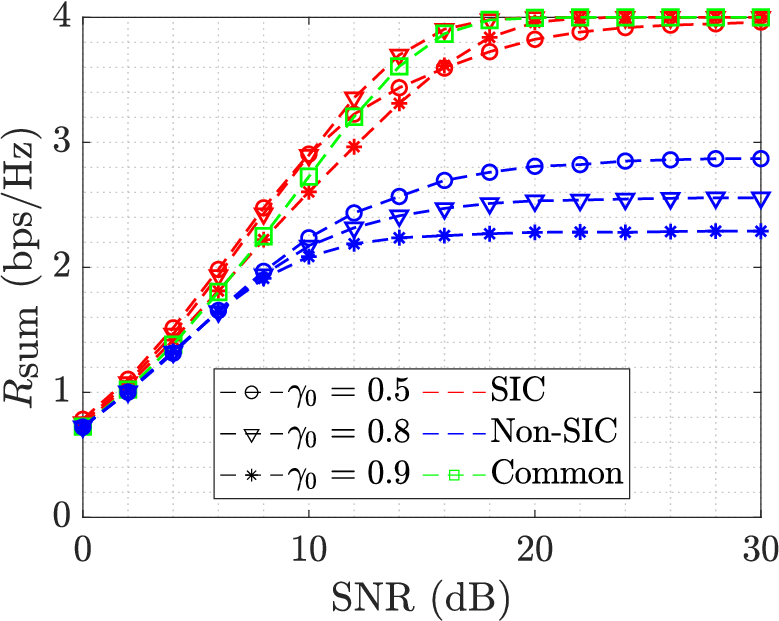}}
\caption{Achievable rate versus SNR for different values of $\textrm{SXPD}_i$. The constellations used are $4$-QAM for the common stream and BPSK ($2$-PAM) for the private streams, with one of them rotated by $90$ degrees. For both the SIC and non-SIC receivers, results are shown for different values of $\gamma_0$. The green line (Common) represents the strategy where the private streams are disabled by setting $\gamma_0 = 1$, and a single $16$-QAM constellation is used.}
\label{fr1}
\end{figure*}

\begin{figure*}
\centering
\subfloat[\normalsize $\textrm{SXPD}_1=\textrm{SXPD}_2=20 \; \textrm{dB},\; \gamma_1=\gamma_2$ \label{fr2_1}]{%
      \includegraphics[width=0.33\linewidth]{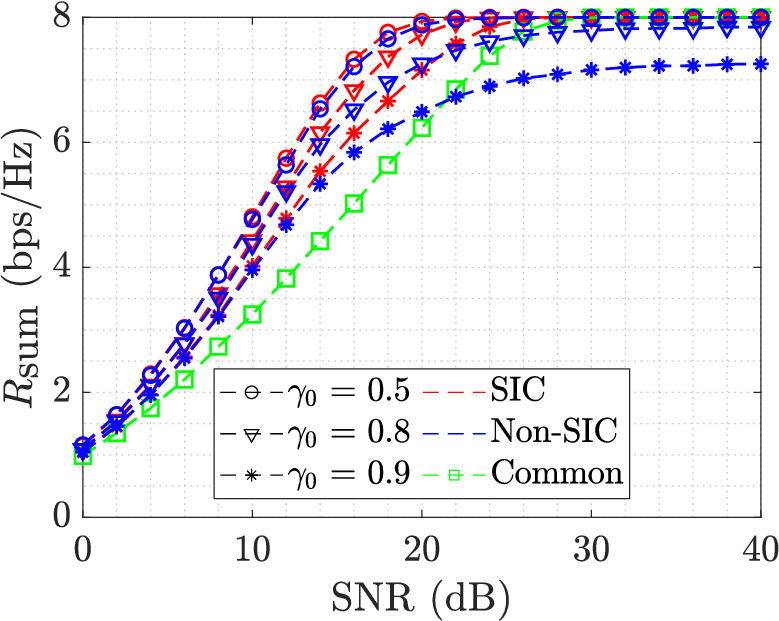}}
      \hfill
\subfloat[\normalsize $\textrm{SXPD}_1=\textrm{SXPD}_2=10 \; \textrm{dB},\; \gamma_1=\gamma_2$\label{fr2_2}]{%
      \includegraphics[width=0.33\linewidth]{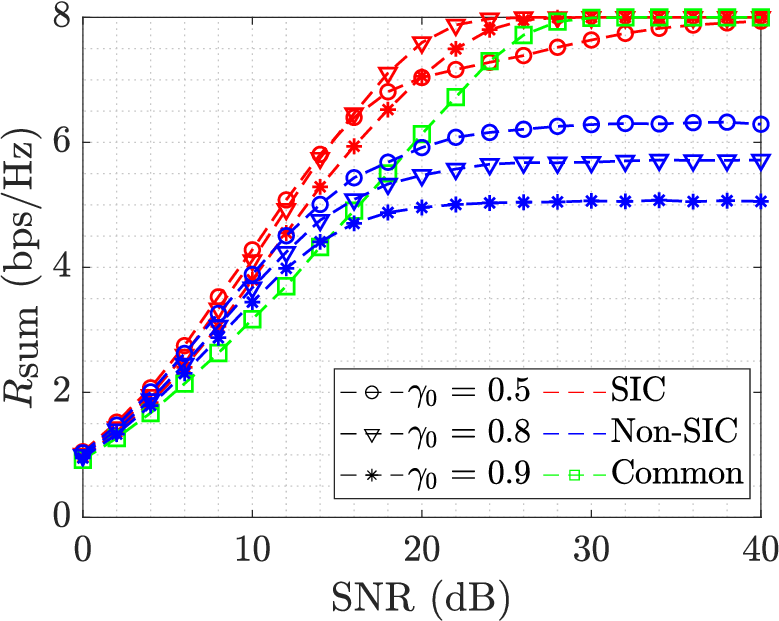}}
        \hfill
\subfloat[\normalsize $\textrm{SXPD}_1=\textrm{SXPD}_2=3 \; \textrm{dB},\; \gamma_1=\gamma_2$\label{fr2_3}]{%
      \includegraphics[width=0.33\linewidth]{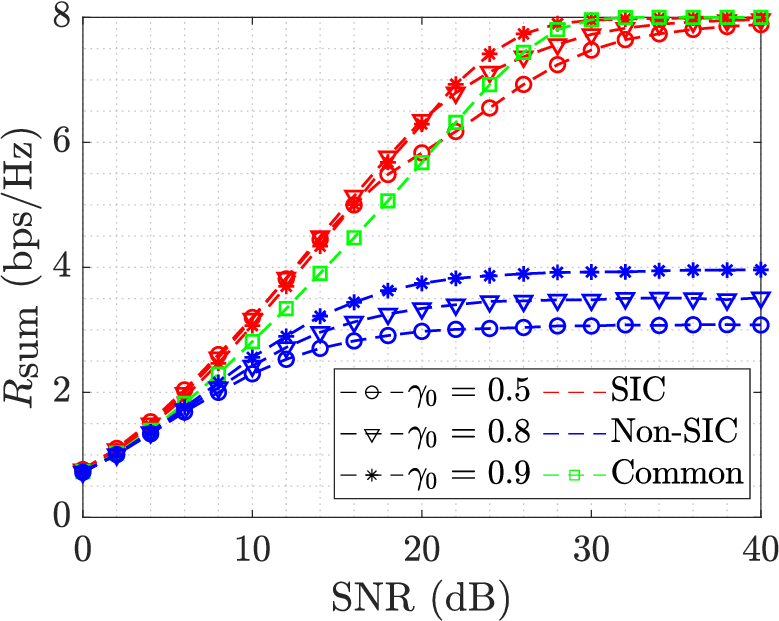}}
\caption{Achievable total rate versus SNR for different values of $\textrm{SXPD}_i$. The constellations used are $16$-QAM for the common stream and $4$-PAM for the private streams, with one of them rotated by $90$ degrees. For both the SIC and non-SIC receivers, results are shown for different values of $\gamma_0$. The green line (Common) represents the strategy where the private streams are disabled by setting $\gamma_0 = 1$, and a single $256$-QAM constellation is used.}
\label{fr2}
\end{figure*}

Fig. \ref{fr1} and Fig. \ref{fr2} show the obtained $R_{\textrm{sum}}$ results for different values of SXPD. For power allocation of the private streams, $\gamma_1=\gamma_2$ is considered, and different values of $\gamma_0$ are depicted for the SIC and non-SIC receivers explained in Section \ref{DRSMA}. In Fig. \ref{fr1}, the common stream uses a $4$-QAM constellation, while each private stream uses a BPSK (or 2-PAM) constellation, with one of them rotated by $90$ degrees. Note that with $\gamma_1=\gamma_2$ the superposition of two PAM constellations in phase quadrature results in a QAM constellation for the superposed alphabet $\mathcal{W}$, and it has been previously used in interference alignment \cite{sisole} and RSMA \cite{rsmapam}. In Fig. \ref{fr2}, the common stream uses a $16$-QAM constellation while the private streams use $4$-PAM constellations. The green line (Common) in both Fig. \ref{fr1} and Fig. \ref{fr2} represents the strategy in which the private streams are disabled by setting $\gamma_0 = 1$. In this case, the common stream uses a $16$-QAM constellation in Fig. \ref{fr1} and a $256$-QAM constellation in Fig. \ref{fr2}.

From Fig. \ref{fr1} and Fig. \ref{fr2}, it can be observed that at a high level of SXPD, with $\textrm{SXPD}_1 = \textrm{SXPD}_2 = 20$ dB, both receivers favor an equal power allocation, corresponding to $\gamma_0 = 0.5$ and $\gamma_1 = \gamma_2 = 0.25$. Note that the constellations used for the private streams have lower cardinality than that of the common stream and therefore require less power for equal performance. Additionally, both receivers exhibit similar performance and achieve a significant improvement over the strategy of using only a single stream (Common). However, as the SXPD decreases, with $\textrm{SXPD}_1=\textrm{SXPD}_2=10$ dB and  $\textrm{SXPD}_1=\textrm{SXPD}_2=3$, the SIC receiver favors allocating more power to the common stream to ensure reliable decoding in the presence of interference from the private streams. Note that, since perfect SIC is assumed, once the common stream is decoded, the private streams are free from interference caused by the common stream. This is not the case for the non-SIC receiver, which leads to its performance degradation. However, as polarization diversity becomes less exploitable (lower SXPD values), the single-stream (Common) strategy, which does not require SIC and can therefore be easily integrated with the non-SIC receiver architecture, begins to exhibit performance comparable to that of the SIC receiver.

\begin{figure}
\centering
\includegraphics[width=0.7\linewidth]{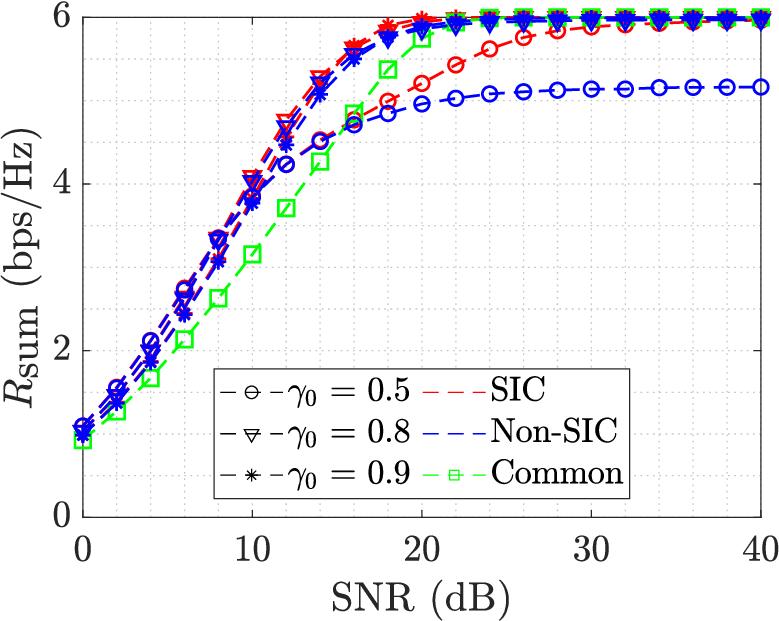}
\caption{Achievable rate versus SNR for $\textrm{SXPD}_1 = 20$ dB and $\textrm{SXPD}_2 = 10$, $\gamma_2 = 0$. The constellations used are $16$-QAM for the common stream and $4$-QAM for the private stream of user $1$. For both the SIC and non-SIC receivers, results are shown for different values of $\gamma_0$. The green line (Common) represents the strategy where both private streams are disabled by setting $\gamma_0 = 1$, and a single $64$-QAM constellation is used.}
\label{fr3}
\end{figure}

In Fig. \ref{fr3}, a scenario with different SXPD values for each user is examined, with $\textrm{SXPD}_1 = 20$ dB and $\textrm{SXPD}_2 = 10$ dB. In this case, the flexibility of RSMA enables disabling one of the private streams by setting $\gamma_2 = 0$, while still allowing the private stream of user $1$ to benefit from polarization diversity. The constellations used are $16$-QAM for the common stream, $4$-QAM for the private stream of user 1, and $64$-QAM for the single-stream (Common) strategy. Once again, SIC and non-SIC receivers exhibit similar performance with appropriate power allocation. Finally, note that, with RSMA, user fairness in terms of rate can still be achieved by adjusting the allocation factor $\beta_i$, even when one private stream is turned off.

\subsection{Numerical Results and System Design, Different Channels}\label{NR_DC}

\begin{figure}
\centering
 \subfloat[\normalsize $\textrm{SXPD}_1=\textrm{SXPD}_2=20 \; \textrm{dB},\gamma_2=0.9(1-\gamma_0)$\label{fr4_1}]{%
       \includegraphics[width=0.7\linewidth]{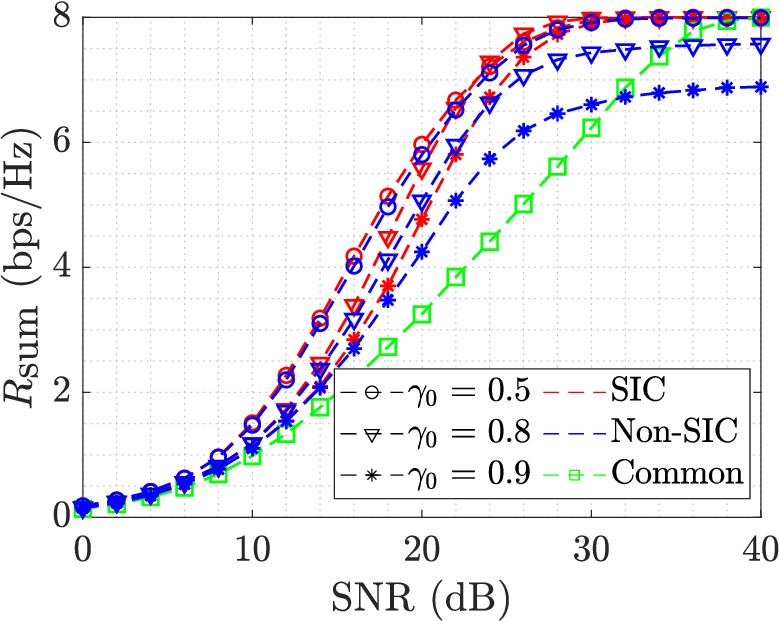}}
  \hfill
  \subfloat[\normalsize $\textrm{SXPD}_1=\textrm{SXPD}_2=3  \; \textrm{dB}$ \label{fr4_2}]{%
        \includegraphics[width=0.7\linewidth]{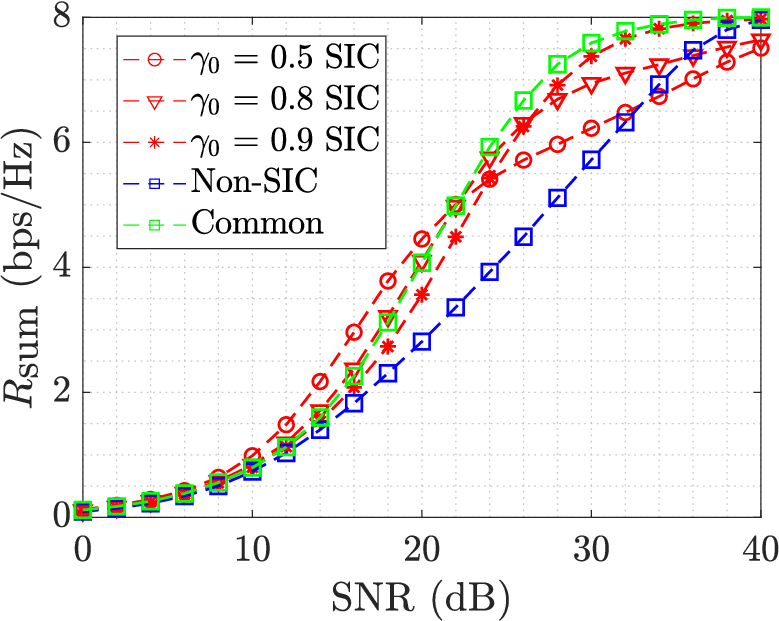}}
  \hfill
  \caption{Total rate versus SNR for different channel strengths. In (a), the constellations used are $16$-QAM for the common stream and $4$-QAM for the private streams. For both the SIC and non-SIC receivers, results are shown for different values of $\gamma_0$. In (b), the non-SIC strategy disables the common stream with $\gamma_0=0$, uses $16$-QAM for both private streams, and sets $\gamma_2=0.95$. The green line (Common) represents the strategy where the private streams are disabled by setting $\gamma_0 = 1$, and a single $256$-QAM constellation is used.}
  \label{fr4}
\end{figure}

We now consider a scenario where the users experience different channel conditions, with $\left|r_{1\textrm{PL}}\right| = 1$ and $\left|r_{2\textrm{PL}}\right| = 0.3162$, resulting in a channel strength difference of $20\log_{10}\left(\frac{\left|r_{1\textrm{PL}}\right|}{\left|r_{2\textrm{PL}}\right|}\right) \approx 10$ dB. Fig. \ref{fr4} presents the results obtained for different values of SXPD and strategies. The disparity in channel strength enables the superposition of private streams in the power domain while preserving user fairness, an approach that represents the core principle of power-domain NOMA \cite{Mod_Cod_NOMA}, and which is incorporated within the RSMA framework. In Fig. \ref{fr4_1}, the constellations used are $16$-QAM for the common stream and $4$-QAM for the private streams. The power allocation is set such that $\gamma_2 = 0.9(1 - \gamma_0)$, leading to $\gamma_2 = 9\gamma_1$, and $\gamma_1\left|r_{1\textrm{PL}}\right|^2 \approx \gamma_2\left|r_{2\textrm{PL}}\right|^2$.  As a result, the trends observed are similar to those in Fig. \ref{fr2_1}, with a rightward shift caused by the degradation in one of the channels. In Fig. \ref{fr4_2}, the scenario with poor SXPD is examined, with $\textrm{SXPD}_1 = \textrm{SXPD}_2 = 3$ dB. In this case, to manage inter-stream interference, the non-SIC strategy disables the common stream by setting $\gamma_0 = 0$, uses $16$-QAM for both private streams, and sets $\gamma_2 = 0.95$. Remarkably, by exploiting channel strength diversity, the non-SIC (combined with the Common strategy) receiver achieves performance close to that of the SIC receiver, even as SXPD decreases.

\subsection{Receiver Selection}\label{NE}

Across all scenarios depicted in Fig. \ref{fr1}, Fig. \ref{fr2}, Fig. \ref{fr3}, and Fig. \ref{fr4}, the combination of the non-SIC receiver with the single-stream (Common) strategy achieves performance that nearly matches that of SIC, while relying on a simpler receiver architecture. The only exception is observed in Fig. \ref{fr2_2} for moderate SXPD, where, compared to Fig. \ref{fr1_2}, constellations with higher cardinality (and thus more susceptible to interference) are considered. Furthermore, it is important to note that the theoretical conditions considered in this study implicitly favor the performance of the SIC receiver, as the assumption of infinite blocklength enables perfect SIC. In contrast, for the non-SIC receiver with partial channel state information at the receiver (CSIR), inter-stream interference is modeled in a worst-case manner as AWGN (see Appendix \ref{appa}). This implies that, in practical implementations, the SIC receiver is likely to experience performance degradation compared to this analysis, whereas the non-SIC receiver is expected to maintain performance more closely aligned with the theoretical findings. In light of these results and considerations, the practical implementation presented in Section \ref{PI} is based on a non-SIC architecture. Results based on SER are provided to demonstrate that the theoretical and experimental outcomes are well aligned. Nonetheless, it must be noted that these conclusions are specific to the chosen RSMA design and the scenario considered, which includes only one pair of dual-polarized transmit antennas and two users. While we comment on the scalability of our proposal in Section \ref{SCA}, \cite{dualRSMA} provides more general insights into SIC and non-SIC architectures for the massive MIMO (e.g. $50$ pairs of dual-polarized antennas) case with dual-polarized RSMA.

\subsection{Magnitude and Polarization CSIT Considerations}\label{ICSIT}

\begin{figure}
\centering
\includegraphics[width=0.7\linewidth]{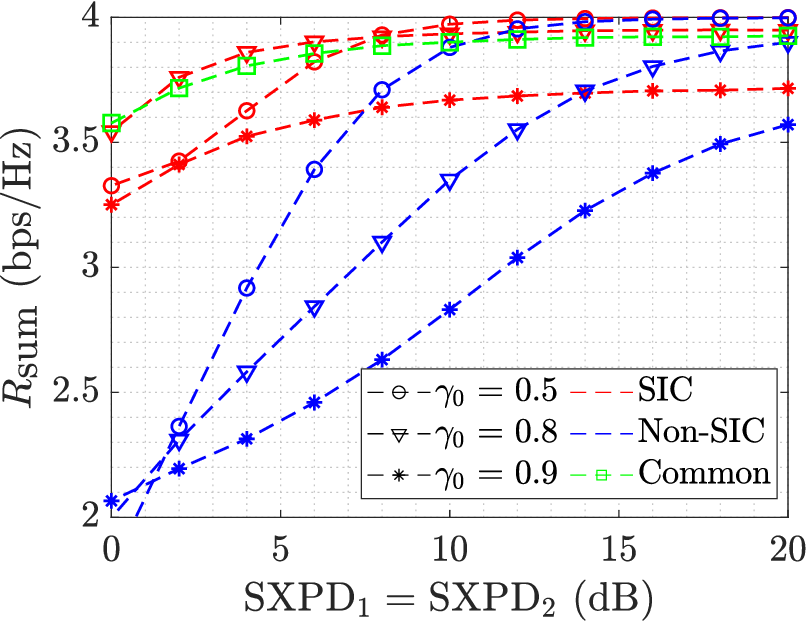}
\caption{Achievable rate versus $\textrm{SXPD}_1 = \textrm{SXPD}_2$ with $\textrm{SNR} = 15$ dB for the same constellation and power allocation values of Fig. \ref{fr1}.}
\label{fricsit}
\end{figure}

Although accurate and updated magnitude and polarization CSIT are more feasible to obtain than phase CSIT in both mmWave bands and SatCom \cite{csitm} due to their slower temporal variation under LoS conditions (Section \ref{CM}), their standard acquisition based on feedback information sent by the receiver will always result in some level of imperfect CSIT because of effects such as pilot reuse, estimation errors, pilot contamination, limited and quantized feedback, and feedback delays. Nonetheless, the consideration of imperfect CSIT motivates conservative approaches to power allocation and constellation design that align well with the frameworks proposed in Section \ref{NR_SC} and Section \ref{NR_DC}. First, the superposed constellation design, specifically the choice between QAM and PAM constellations, is primarily based on whether there is a significant disparity in channel strength across users \cite{rsmapam}, which can generally be determined at the transmitter even with imperfect CSIT. Second, the number of symbols assigned to both the common and private streams and power allocation are selected based on the SNR and polarization conditions, where design margins can be applied to enhance robustness. Regarding this last point, it must be noted that, for given power allocation and SNR values, the system performance tends to vary smoothly with the polarization conditions, with the total rate increasing with $\textrm{SXPD}_1$ and $\textrm{SXPD}_2$ up to the upper bound imposed by the chosen constellation. To illustrate this, Fig. \ref{fricsit} shows achievable rate results for the same constellation and power allocation values as in Fig. \ref{fr1}, in this case fixing the SNR at $15$ dB and considering different values of $\textrm{SXPD}_i$ with $\left|r_{1\textrm{PL}}\right| = \left|r_{2\textrm{PL}}\right| = 1$. Although it is clear that, due to the specific constellations chosen, some power allocation values provide better performance, all configurations consistently improve with polarization diversity without abrupt changes, indicating a certain degree of robustness against imperfect knowledge of $\textrm{SXPD}_i$. In other words, the negative impact of selecting a particular power allocation value based on imperfect polarization CSIT is limited.

Overall, these considerations support a constellation and power allocation design based on precomputed look-up tables, as commonly used in wireless standards that implement adaptive modulation and coding, such as 4G and 5G. Of course, this solution does not guarantee optimality for specific channel conditions, but it offers a reasonable trade-off between exploiting these conditions, system stability, and robustness against imperfect CSIT. Furthermore, these look-up tables can be adapted depending on the application and the degree of CSIT uncertainty. Naturally, when phase CSIT is available, these look-up tables can be extended to include indices of precoding vectors \cite{codebook_tutorial}. Finally, it must be noted that, although beyond the scope of this paper, the joint optimization of RSMA precoding and constellation design, considering both discrete constellations and imperfect CSIT, is a promising direction for future research aimed at further improving performance without compromising practicality \cite{ojcom}.

\subsection{Comparison with RSMA Full Polarization Diversity}\label{Comp}

Before addressing the scalability of our design, we compare our proposal with a scheme that transmits replicas of all streams over both polarizations in the scenario considered. This scheme, also proposed in \cite{dualRSMA} for GS, can be implemented for DS by setting $x_{\textrm{V}} = x_{\textrm{H}} = \frac{1}{\sqrt{2}}(x_0 + x_1 + x_2) = \frac{1}{\sqrt{2}}(x_0 + w)$. The factor $\frac{1}{\sqrt{2}}$ is introduced to preserve the power allocation notation while equally splitting the transmitted power between the two polarizations. We define the combined received signal, $y$, as: 

\begin{equation}\label{fdp}
\begin{aligned}
&y = y_{i\textrm{V}} + y_{i\textrm{H}} =& \\
&\frac{1}{\sqrt{2}} (x_0 + w) (h_{i\textrm{VV}} + h_{i\textrm{HH}} + h_{i\textrm{HV}} + h_{i\textrm{VH}}) + n_{i\textrm{V}} + n_{i\textrm{H}}, \\
\end{aligned}
\end{equation} where some degree of polarization diversity is achieved for every stream. Considering only SIC decoding, the receiver follows the same strategy described in Section \ref{SIC}: the common stream is decoded first, SIC is then performed, and the superposition of private streams is subsequently decoded using JMLD. With this in mind, it is straightforward to adapt the mutual information expressions provided in Appendix \ref{appa} by appropriately modifying the channel coefficients.

\begin{figure}
\centering
\includegraphics[width=0.7\linewidth]{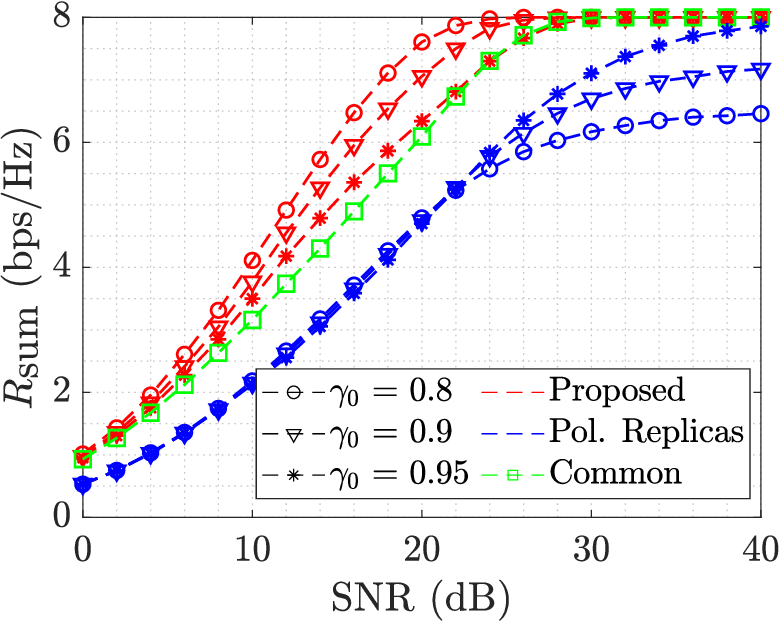}
\caption{Achievable rate versus SNR for $\textrm{SXPD}_1 = \textrm{SXPD}_2 = 10$ dB, using the same constellation and power allocation values as in Fig. \ref{fr2_2}. In both cases, SIC decoders are applied.}
\label{frfd}
\end{figure}

However, a limitation of this design in the scenario considered is that, because only a single pair of dual-polarized antennas is used for transmission, separate precoding cannot be applied to the common and private streams, even when phase CSIT is available. Furthermore, the channel coefficients in Eq. (\ref{fdp}) do not generally combine coherently, whereas the summation of the noise components duplicates the variance of the AWGN. To illustrate this effect, Fig. \ref{frfd} shows the achievable rates for $\textrm{SXPD}_1 = \textrm{SXPD}_2 = 10$ dB, using the same constellation and power allocation parameters as in Fig. \ref{fr2_2}. SIC decoding is applied in both cases, and the scheme in which all streams are transmitted over each polarization is denoted by Pol. Replicas. It can be observed that the proposed design achieves higher rates even at moderate cross-polarization rejection levels, and that, in this scenario, allocating all available power to the common stream also yields superior performance. This is because, without additional antenna precoding, inter-stream interference is maximized in the Pol. Replicas scheme, which consequently requires higher values of $\gamma_0$ (e.g., $\gamma_0 = 0.95$ in Fig. \ref{frfd}) to ensure reliable SIC.

\subsection{Scalability to More Users}\label{SCA}

\begin{figure}
\centering
\includegraphics[width=0.7\linewidth]{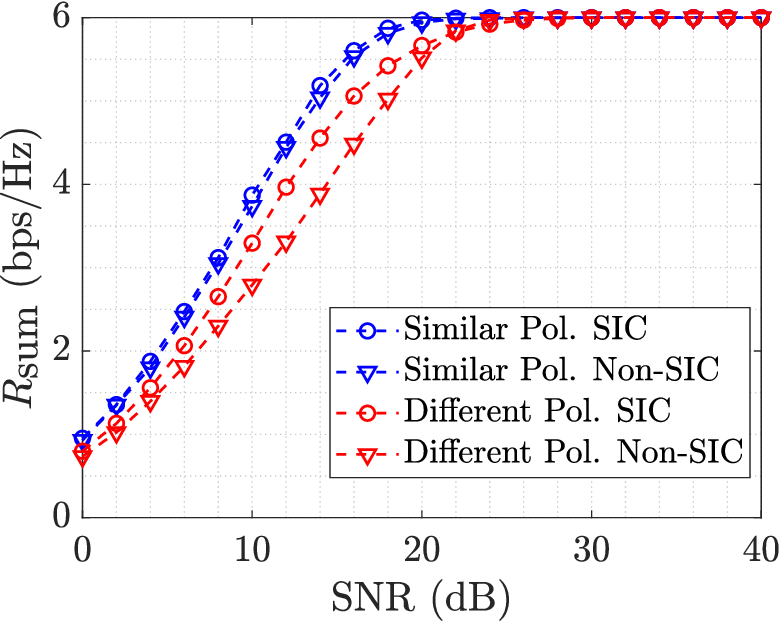}
\caption{Average achievable rate versus SNR for $\textrm{SXPD}_1 = \textrm{SXPD}_2 = 20$ dB and $\textrm{SXPD}_3 = \textrm{SXPD}_4 = 3$ dB, comparing the pairing of users with similar and different polarization conditions (Similar Pol. and Different Pol., respectively).}
\label{froth}
\end{figure}

Finally, we briefly discuss the scalability of the system to a larger number of users. Two scenarios can be considered: increasing the number of users together with the number of available antennas and increasing only the number of users. In the first scenario, phase CSIT and phase coordination across transmit antennas are generally required to efficiently multiplex users in the spatial domain, in addition to exploiting the power, code, and polarization domains. The potential and performance limits of dual-polarized RSMA with an increasing number of antennas, including massive MIMO configurations, have already been studied in \cite{dualRSMA}. Imperfect CSIT in dual-polarized RSMA remains relatively unexplored, although the intrinsic robustness of RSMA to CSIT imperfections is expected to extend to this scenario.

More specific to our contribution is the case in which the number of users increases while each transmitter and receiver is still equipped with a single pair of dual-polarized antennas. This configuration is of practical relevance in mmWave and SatCom scenarios because of hardware constraints. In such cases, the proposed technique can always be combined with orthogonal multiplexing in the time, frequency, or code domains. For example, four users can be served over two orthogonal frequency bands, using RSMA to serve two users, rather than one, in each band. Although this is a straightforward and robust approach, the flexibility of RSMA also enables a more efficient use of resources under certain channel conditions. For instance, three users with significant disparities in channel strength can be multiplexed in the power domain by keeping a single common stream and adding an additional private stream. However, the efficient design of the superposed constellation would depend on the aforementioned channel-strength disparity \cite{ojcom}.

Although pairing users with different channel strengths is beneficial for power-domain multiplexing, we preliminarily investigate, through an illustrative example, whether the same principle applies to polarization conditions. We consider four users with $\textrm{SXPD}_1=\textrm{SXPD}_2 = 20$ dB and $\textrm{SXPD}_3=\textrm{SXPD}_4 = 3$ dB under identical channel-strength conditions, i.e., $\left|r_{1\textrm{PL}}\right| = \left|r_{2\textrm{PL}}\right| = \left|r_{3\textrm{PL}}\right| = \left|r_{4\textrm{PL}}\right| = 1$. Two pairing strategies are considered. The first pairs users with similar polarization conditions within each resource block, i.e., users $1$ and $2$ are paired and served through RSMA, while being orthogonal in time, frequency, or code to the pair of users $3$ and $4$. The second strategy pairs users with different polarization conditions, e.g., users $1$ and $3$, and users $2$ and $4$. Fig. \ref{froth} shows the achievable-rate results for SIC and non-SIC decoding under both pairing strategies, denoted as Similar Pol. and Different Pol., respectively. When users $1$ and $2$ are paired, the same constellations as in Fig. \ref{fr2_1} are used, and the best-performing power allocation value observed for both decoders, $\gamma_0=0.5$, is selected. When users $3$ and $4$ are paired, the constellations are the same as those in Fig. \ref{fr1_3}: the non-SIC decoder allocates all the power to the common stream, whereas the SIC decoder uses $\gamma_0=0.8$. When either users $1$ and $3$, or users $2$ and $4$ are paired, the constellations are the same as those in Fig. \ref{fr3}: the non-SIC decoder also allocates all the power to the common stream, while the SIC decoder uses $\gamma_0=0.9$. The user pairs are served orthogonally with an equal allocation of resources; therefore, the average achievable rate is computed as the arithmetic mean of the $R_{\textrm{sum}}$ for each pair. Interestingly, the results show that, in this scenario, pairing users with similar polarization conditions achieves higher rates when the constellations and power-allocation parameters are adjusted accordingly. The main reason is that polarization diversity is fully exploited when both users exhibit high levels of cross-polarization rejection. When this condition is satisfied by only one of the two users, polarization diversity still provides some gains, as shown in Fig. \ref{fr3}, although these gains are more limited.

\section{Practical Implementation}\label{PI}

In this section, we present the prototype implementation based on SDR for the experimental validation of the system. The results obtained using this prototype are presented in Section \ref{ER}.

\subsection{General Considerations}\label{gen_impl}
Baseband digital signal processing is designed and implemented using the GNU Radio framework, an open-source software platform based on C++ and Python, which can be executed on any generic CPU, such as a personal computer (PC). Universal Software Radio Peripherals (USRPs), model USRP X310 with two UBX 10-6000 daughterboards each, are used to serve as the transmitter and receiver, performing analog-to-digital and digital-to-analog conversion (ADC/DAC) as well as intermediate frequency (IF) conversion. Since USRP devices are compatible with GNU Radio, communication parameters such as gain, carrier frequency, sampling rate, constellation, and power allocation can be adjusted via software within the limits supported by the hardware. As the carrier frequency of USRP devices is limited to a maximum of $6$ GHz, commercial mixers and power amplifiers (models Mini-Circuits ZMDB-44H-K+ and Marki Microwave AMM-6702UC, respectively) are added to the front end of each device to achieve the chosen carrier frequency of $25$ GHz, while the IF is set to $2.5$ GHz. For the antennas, we use the commercial dual-linearly polarized horn antenna model ERAVANT SAF-2434031535-358-S1-280-DP. Each antenna has two independent ports, which are connected to different ports of the USRP. From a theoretical perspective, this setup is equivalent to having two single-polarized antennas, each connected to a separate port.

The communication architecture and packet structure is based on our previous work \cite{tcom}, using a single-carrier configuration and transmitting small packets (a few hundred bytes) composed of three fields: first, an access code of 8 bytes to identify the beginning of the packet; second, a cyclic redundancy check 32 (CRC32) sequence of 4 bytes for error detection; and third, a payload of variable length (default 200 bytes). The estimation of SER based on the CRC32 field of the received packets is carried out as described in \cite[Eq.~(14)]{tcom} and \cite{elec}. The sampling rate of each port is set to $10$ Msps, using two digital samples per symbol and a roll-off factor of $0.35$ in the root-raised cosine (RRC) filter, resulting in a communications channel bandwidth of $6.75$ MHz. Note that the sampling rate is severely limited by the over-the-air processing capabilities of the USRP devices connected to a laptop. As a result, the prototype can be viewed as testing a single subchannel of a potential mmWave system. Therefore, spectral efficiency and SER are chosen as figures of merit to evaluate performance rather than total throughput. Potential integration with OFDM and/or SC-FDMA architectures, which has already been examined in theoretical studies \cite{rsma_ofdm}, is left for future work.

Note that the choice of the SER, rather than the bit error rate (BER), as the error metric is motivated by three aspects. First, when transmitting random bits (as in our case), BER estimation is more complex and potentially less accurate than SER estimation when both are inferred from evaluating the CRC32 header appended to the packets. Second, the use of SER as a performance metric is particularly relevant in scenarios involving superposition coding, where inter-user interference typically constitutes the main performance-limiting factor. Third, SER enables a direct comparison between the experimental measurements and theoretical Monte Carlo simulations for uncoded communication (i.e., without redundancy for error correction). Although beyond the scope of this paper, joint bit labeling and channel coding strategies explicitly designed for error detection and correction in RSMA constitute a promising yet largely unexplored research direction in the literature \cite{ojcom,Mod_Cod_NOMA}.

\begin{figure*}[t]
\centering
\noindent
\begin{minipage}[t]{0.48\textwidth}
\centering
\resizebox{\linewidth}{!}{%
\begin{tikzpicture}

  \node[draw, rectangle, align=center, fill=blue!30] (block1) {User $1$\\Bits};

  \node[draw, rectangle, align=center, below=1cm of block1, fill=blue!30] (block1b) {User $2$\\Bits};

  \node[draw, rectangle, fill=blue!30, align=center, right = 0.5cm of block1, yshift=-1cm,
    minimum height=3.5cm, minimum width=1.0cm
  ] (splitter) {\rotatebox{90}{Message Splitter}};

  \node[draw, rectangle, align=center, right=0.5cm of splitter, yshift=+1.3cm, fill=blue!30] (block3) {Source\\Coding ($\mathcal{X}_0$)};
  \node[draw, rectangle, align=center, right=0.5cm of splitter, yshift=-0cm, fill=blue!30] (block3b) {Source\\Coding ($\mathcal{X}_1$)};
   \node[draw, rectangle, align=center, right=0.5cm of splitter, yshift=-1.3cm, fill=blue!30] (block3c) {Source\\Coding ($\mathcal{X}_2$)};

  \draw[->, line width=0.7mm] (block1.east) -- (splitter.west |- block1);
  \draw[->, line width=0.7mm] (block1b.east) -- (splitter.west |- block1b);
  \draw[->, line width=0.7mm] (splitter.east|- block3) -- (block3.west);
  \draw[->, line width=0.7mm] (splitter.east |- block3b) -- (block3b.west);
  \draw[->, line width=0.7mm] (splitter.east |- block3c) -- (block3c.west);

  \draw[->, line width=0.7mm] (block3b.east) -- ++(0.8,0) -- ++(0,-0.25) node[midway, above=3mm] {\fontsize{12}{14}\selectfont $x_{1m}$};
  \draw[->, line width=0.7mm] (block3c.east) -- ++(0.8,0) -- ++(0,0.25) node[midway, below=3mm] {\fontsize{12}{14}\selectfont $x_{2l}$};

  \node[draw, line width=0.25mm, circle, align=center, right=0.42cm of block3, yshift=-1.95cm, minimum size=0.5cm, fill=blue!30] (sum) {$\boldsymbol{+}$};

 \node[draw, rectangle, align=center, right=2.15cm of block3, fill=blue!30] (block4a) {Pulse\\Shaping};

   \node[draw, rectangle, align=center, right=0.5 cm of block4a, fill=red!30] (block51) {USRP Tx\\ (Port 1)};

   \node[draw, rectangle, align=center, right=0.5 cm of block51, fill=red!30] (block61) {Mixer \\+ Amplifier};

\draw[->, line width=0.7mm] (block4a) -- (block51);
\draw[->, line width=0.7mm] (block51) -- (block61);

   \draw[line width=0.8mm] (block61.east) -- ++(0.6,0) -- ++(0,0.6)
      node[midway, right, align=left] {Tx\\ V Antenna};

  \node[draw, regular polygon, regular polygon sides=3, shape border rotate=180, minimum size=1cm, line width=0.8mm, anchor=south, fill=red!30] at ($(block61.east) + (0.6,0.6)$) {};

  \node[draw, rectangle, align=center, right=1cm of sum, fill=blue!30] (block4) {Pulse\\Shaping};

  \node[draw, rectangle, align=center, right=0.5cm of block4, fill=red!30] (block5) {USRP Tx \\ (Port 2)};
  \node[draw, rectangle, align=center, right=0.5cm of block5, fill=red!30] (block6) {Mixer \\+ Amplifier};

  \draw[->, line width=0.7mm] (block3) -- (block4a) node[midway, above] {\fontsize{12}{14}\selectfont $x_{0n}$};
  \draw[->, line width=0.7mm] (sum) -- (block4) node[midway, above] {\fontsize{12}{14}\selectfont $w_{ml}$};

  \draw[line width=0.8mm] (block6.east) -- ++(0.6,0) -- ++(0,0.6)
      node[midway, right, align=left] {Tx\\ H Antenna};

  \node[draw, regular polygon, regular polygon sides=3, shape border rotate=180, minimum size=1cm, line width=0.8mm, anchor=south, fill=red!30] at ($(block6.east) + (0.6,0.6)$) {};

\draw[->, line width=0.7mm] (block4) -- (block5);
\draw[->, line width=0.7mm] (block5) -- (block6);

\end{tikzpicture}%
}
\end{minipage}%
\hfill%
\begin{minipage}[t]{0.48\textwidth}
\centering
\raisebox{1mm}{
\resizebox{\linewidth}{!}{%
\begin{tikzpicture}
\node[draw, rectangle, align=center, fill=red!30] (rblock1) {Mixer};
\node[draw, rectangle, align=center, right=0.5cm of rblock1, fill=red!30] (rblock2) {USRP Rx\\ (Port 1)};
\node[draw, rectangle, align=center, right=0.5cm of rblock2, fill=blue!30] (rblock3) {Time Sync. \& \\ Channel \\ Correction ($\mathcal{X}_0$)};
\node[draw, rectangle, align=center, right=0.5cm of rblock3, fill=blue!30] (rblock4) {Symbol \\ Decoding ($\mathcal{X}_0$)};

\node[draw, rectangle, align=center, right=3cm of rblock4, fill=blue!30, yshift=-1.2cm,
      minimum height=3.5cm, minimum width=1.0cm] (rblock5) {\rotatebox{90}{Packet Decoding}};

\draw[->, line width=0.7mm] (rblock1) -- (rblock2);
\draw[->, line width=0.7mm] (rblock2) -- (rblock3);
\draw[->, line width=0.7mm] (rblock3) -- (rblock4);
\draw[->, line width=0.7mm] (rblock4.east) -- (rblock5.west |- rblock4) node[midway, above] {\fontsize{12}{14}\selectfont $\hat{x}_{i0}$};

\draw[line width=0.8mm] (rblock1.west) -- ++(-0.6,0) -- ++(0,0.6);
\node[left=0.1cm of rblock1,align = left] {Rx\\ V Antenna};
\node[draw, regular polygon, regular polygon sides=3, shape border rotate=180, minimum size=1cm,
      line width=0.8mm, anchor=south, fill=red!30]
      at ($(rblock1.west) + (-0.6,0.6)$) {};

\node[draw, rectangle, align=center, below=1.5cm of rblock1, fill=red!30] (rblock1b) {Mixer};
\node[draw, rectangle, align=center, right=0.5cm of rblock1b, fill=red!30] (rblock2b) {USRP Rx\\ (Port 2)};
\node[draw, rectangle, align=center, right=0.5cm of rblock2b, fill=blue!30] (rblock3b) {Time Sync. \& \\ Channel \\ Correction ($\mathcal{W}$)};
\node[draw, rectangle, align=center, right=0.5cm of rblock3b, fill=blue!30] (rblock4b) {JMLD Symbol \\ Decoding ($\mathcal{W}$)};
\node[draw, rectangle, align=center, right=1cm of rblock4b, fill=blue!30] (rblock5b) {Index \\ Selection};

\draw[->, line width=0.7mm] (rblock1b) -- (rblock2b);
\draw[->, line width=0.7mm] (rblock2b) -- (rblock3b);
\draw[->, line width=0.7mm] (rblock3b) -- (rblock4b);
\draw[->, line width=0.7mm] (rblock4b) -- (rblock5b) node[midway, above] {\fontsize{12}{14}\selectfont $\hat{w}_{i}$};

\draw[line width = 0.8mm] (rblock1b.west) -- ++(-0.6,0) -- ++(0,0.6);
\node[left=0.1cm of rblock1b,align=left] {Rx\\ H Antenna};
\node[draw, regular polygon, regular polygon sides=3, shape border rotate=180,
      minimum size=1cm, line width=0.8mm, anchor=south, fill=red!30]
      at ($(rblock1b.west) + (-0.6,0.6)$) {};

\draw[->, line width=0.7mm] (rblock5b.east) -- (rblock5.west |- rblock4b);

\end{tikzpicture}%
}}
\end{minipage}

\caption{Transmitter and receiver diagram for the $i$-th user. Blue and red colors indicate baseband digital signal processing and hardware blocks, respectively.}
\label{txrxschemes}
\end{figure*}
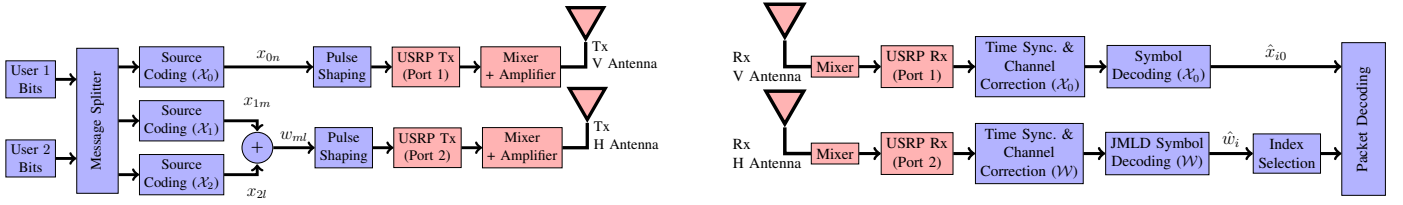

\begin{figure*}
\centering

\subfloat[\label{m1_1}]{%
\begin{tikzpicture}
  \node[inner sep=0] (img) {\includegraphics[width=0.22\linewidth]{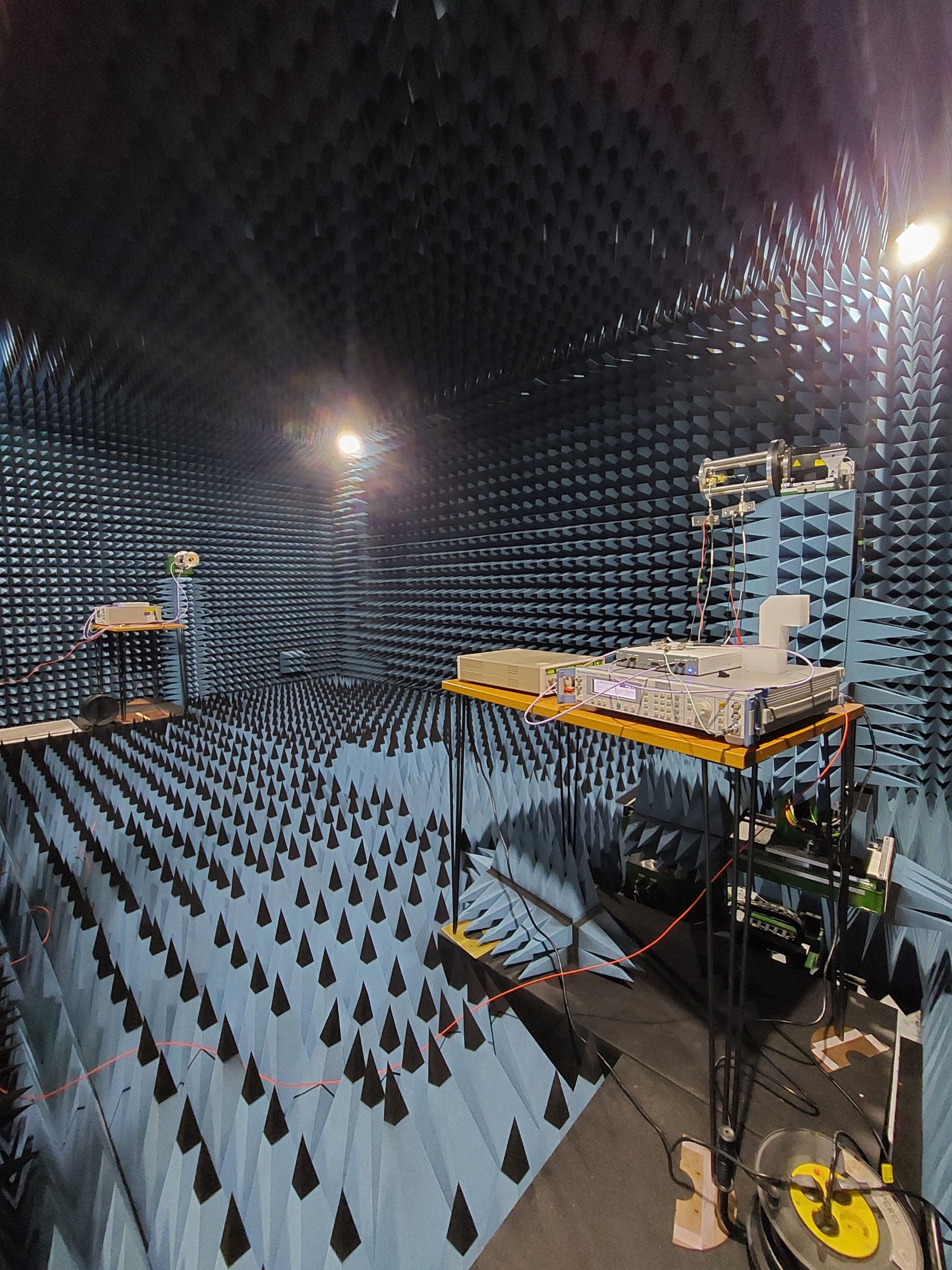}};

\end{tikzpicture}
}
\hfill
\subfloat[\label{m1_2}]{%
\begin{tikzpicture}
  \node[inner sep=0] (img) {\includegraphics[width=0.22\linewidth]{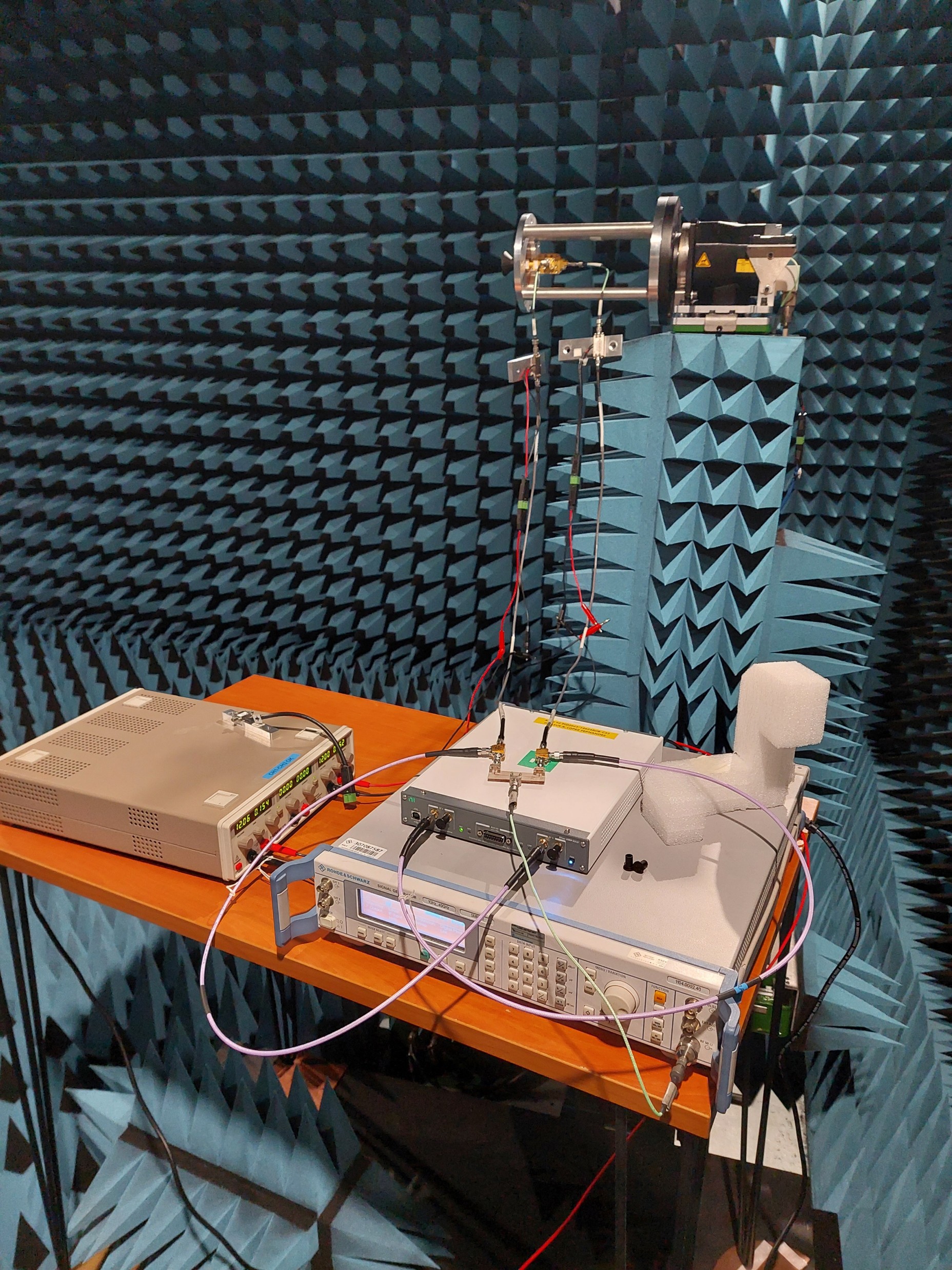}};

  \node[draw, fill=white, rounded corners, font=\scriptsize] 
      at (-0.1, 2) {Tx Antenna};

    \node[
      draw,
      fill=white,
      rounded corners,
      font=\scriptsize,
      align=center
    ] at (-0.55, 1.1) {Power \\ Amplifiers};

   \node[draw, fill=white, rounded corners, font=\scriptsize] 
      at (0.8, -1) {USRP};

    \node[draw, fill=white, rounded corners, font=\scriptsize] 
      at (0.875, -0.25) {Mixers};

    \node[
      draw,
      fill=white,
      rounded corners,
      font=\scriptsize,
      align=center
    ] at (-1.5, -0.4) {Power \\ Supply};

    \node[
      draw,
      fill=white,
      rounded corners,
      font=\scriptsize,
      align=center
    ] at (0, -2) {Local \\ Oscillator};
  
\end{tikzpicture}
}
\hfill
\subfloat[\label{m1_3}]{%
\begin{tikzpicture}
  \node[inner sep=0] (img) {\includegraphics[width=0.22\linewidth]{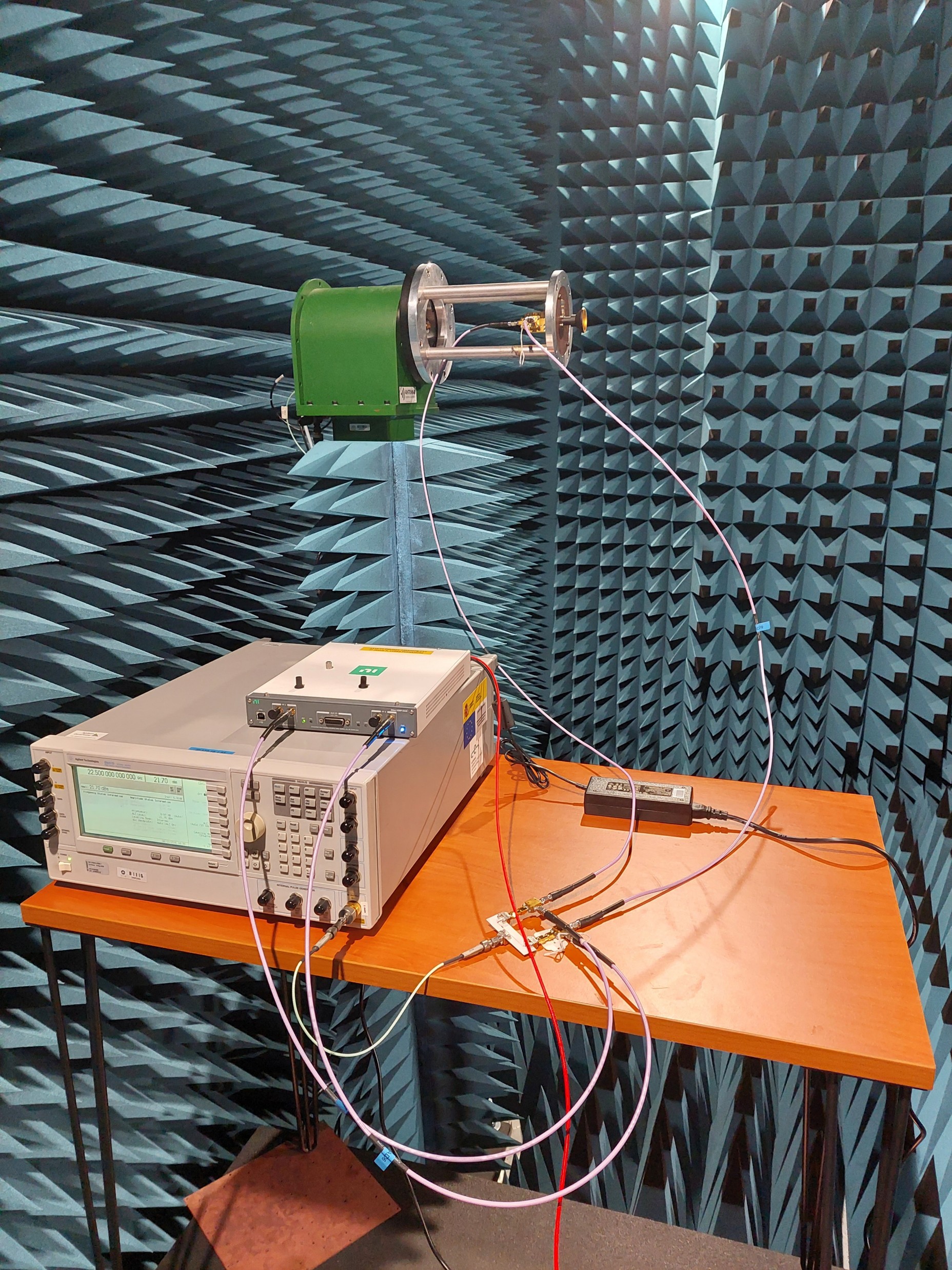}};

   \node[draw, fill=white, rounded corners, font=\scriptsize] 
      at (1.1, 1.7) {Rx Antenna};

   \node[draw, fill=white, rounded corners, font=\scriptsize] 
      at (-1, 0) {USRP};

    \node[draw, fill=white, rounded corners, font=\scriptsize] 
      at (1, -1.3) {Mixers};

    \node[
      draw,
      fill=white,
      rounded corners,
      font=\scriptsize,
      align=center
    ] at (-1.5, -1.5) {Local \\ Oscillator};
  
\end{tikzpicture}
}
\hfill
\subfloat[\label{m1_4}]{%
\begin{tikzpicture}
  \node[inner sep=0] (img) {\includegraphics[width=0.22\linewidth]{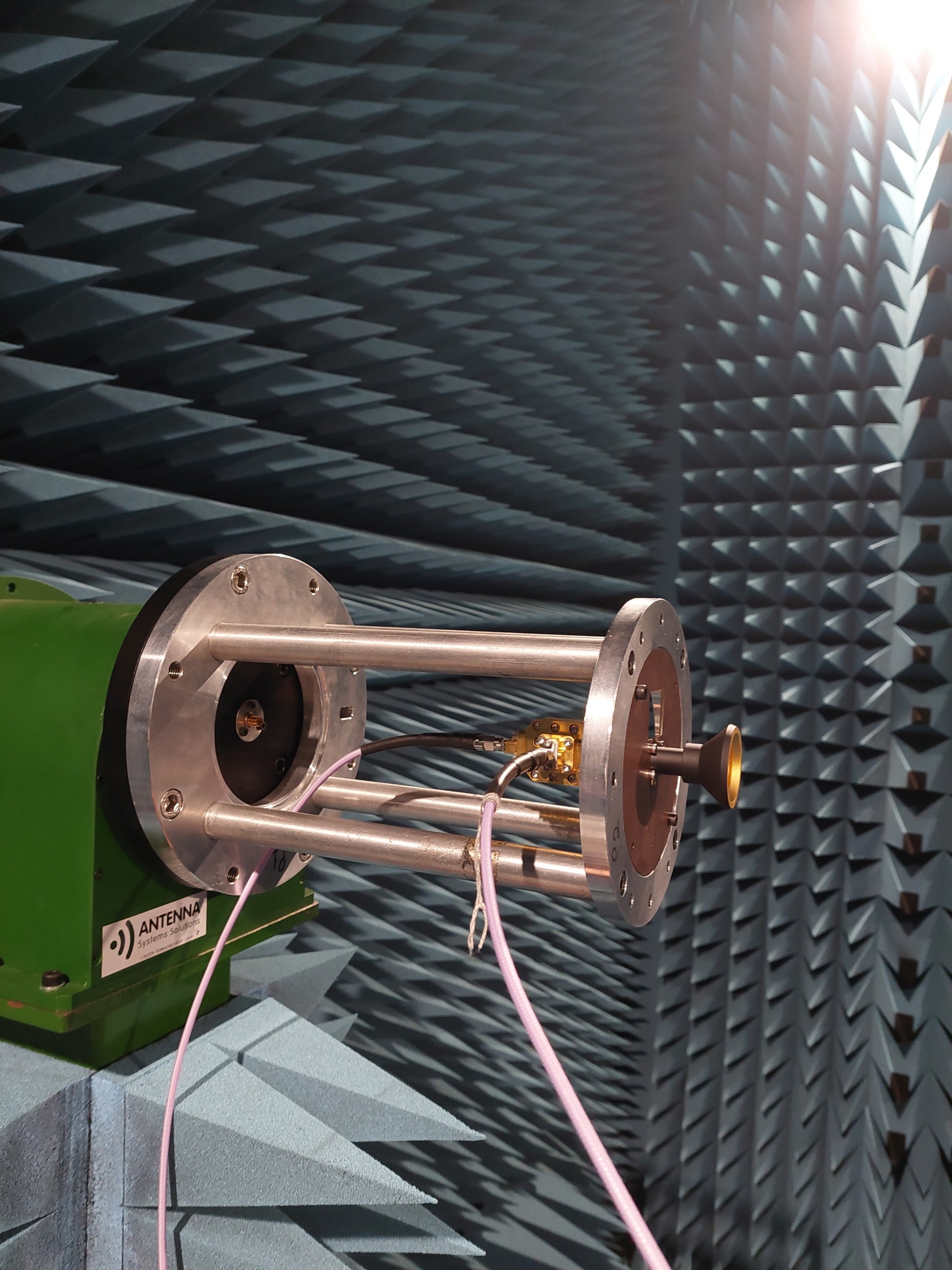}};

   \node[draw, fill=white, rounded corners, font=\scriptsize] 
      at (-0.1,-0.15) {V Pol.};

   \node[draw, fill=white, rounded corners, font=\scriptsize] 
      at (0.55,-0.95) {H Pol.};
      
\end{tikzpicture}
}

\caption{Measurement setup in the anechoic chamber at the University of Oviedo.
(a) Complete setup. (b) Transmitter. (c) Receiver. (d) Dual-polarized antenna.}
\label{photos}
\end{figure*}
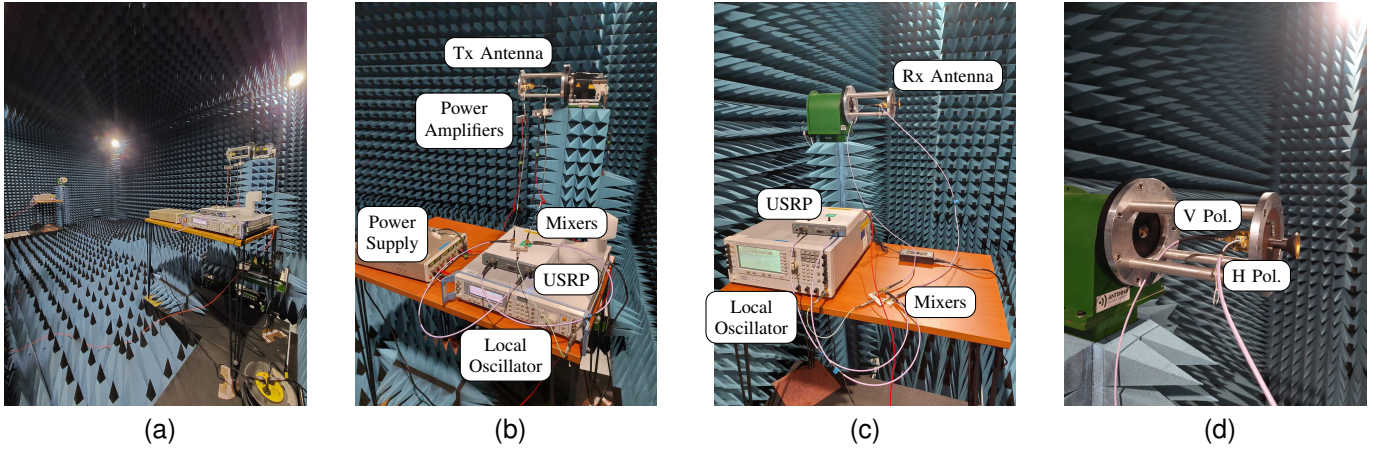

\subsection{Transmitter}\label{tx_impl}
The transmitter diagram is shown in Fig. \ref{txrxschemes}, with blue and red colors indicating digital signal processing and hardware blocks, respectively. The message-splitting block combines the bits from both users to form three data streams and, for each stream, organizes the bits into packets by adding the access code and CRC32 fields (Section \ref{gen_impl}). Note that, although a real-world implementation would generate one packet per user, we choose to generate one packet per stream so that, in Section \ref{ER}, we can evaluate the measured SER of the three streams and gain further insight into the system performance. The source-coding blocks then map the incoming bits to constellation symbols according to the selected constellations and power allocation defined by $\mathcal{X}_k$. Finally, the symbols are resampled and filtered by an RRC filter before being passed to the USRP, which performs DAC and shifts the signal to the IF. Importantly, although power allocation is performed digitally, a normalization is applied such that $\gamma_0 + \gamma_1 + \gamma_2 = 1$, ensuring that the total transmitted power (determined by the hardware power amplifiers) remains constant.

\begin{figure}
\centering
\includegraphics[width=0.95\linewidth]{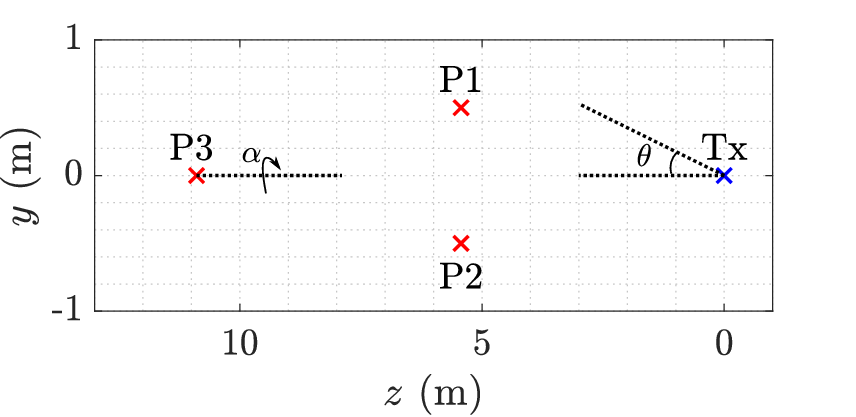}
\caption{Measurement campaign floor plan. Considered positions for the receiver are P1, P2 and P3, while the position of the transmitter is fixed.}
\label{pos}
\end{figure}

\begin{figure*}
\centering
 \subfloat[\normalsize $\chi_\textrm{VV}$ (dB) \label{figvv}]{%
       \includegraphics[width=0.5\linewidth]{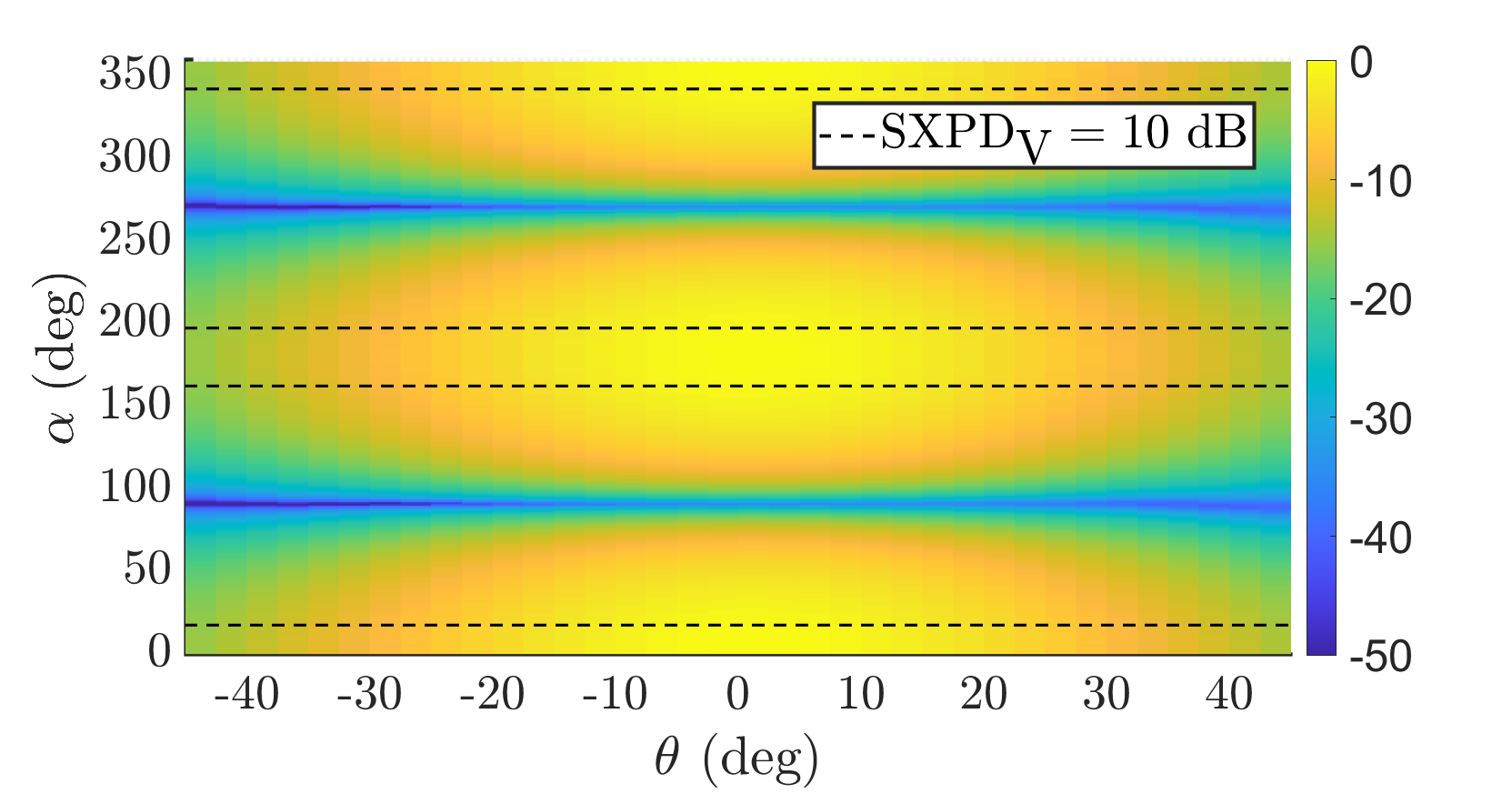}}
  \hfill
  \subfloat[$\chi_\textrm{VH}$ (dB) \label{figvh}]{%
        \includegraphics[width=0.5\linewidth]{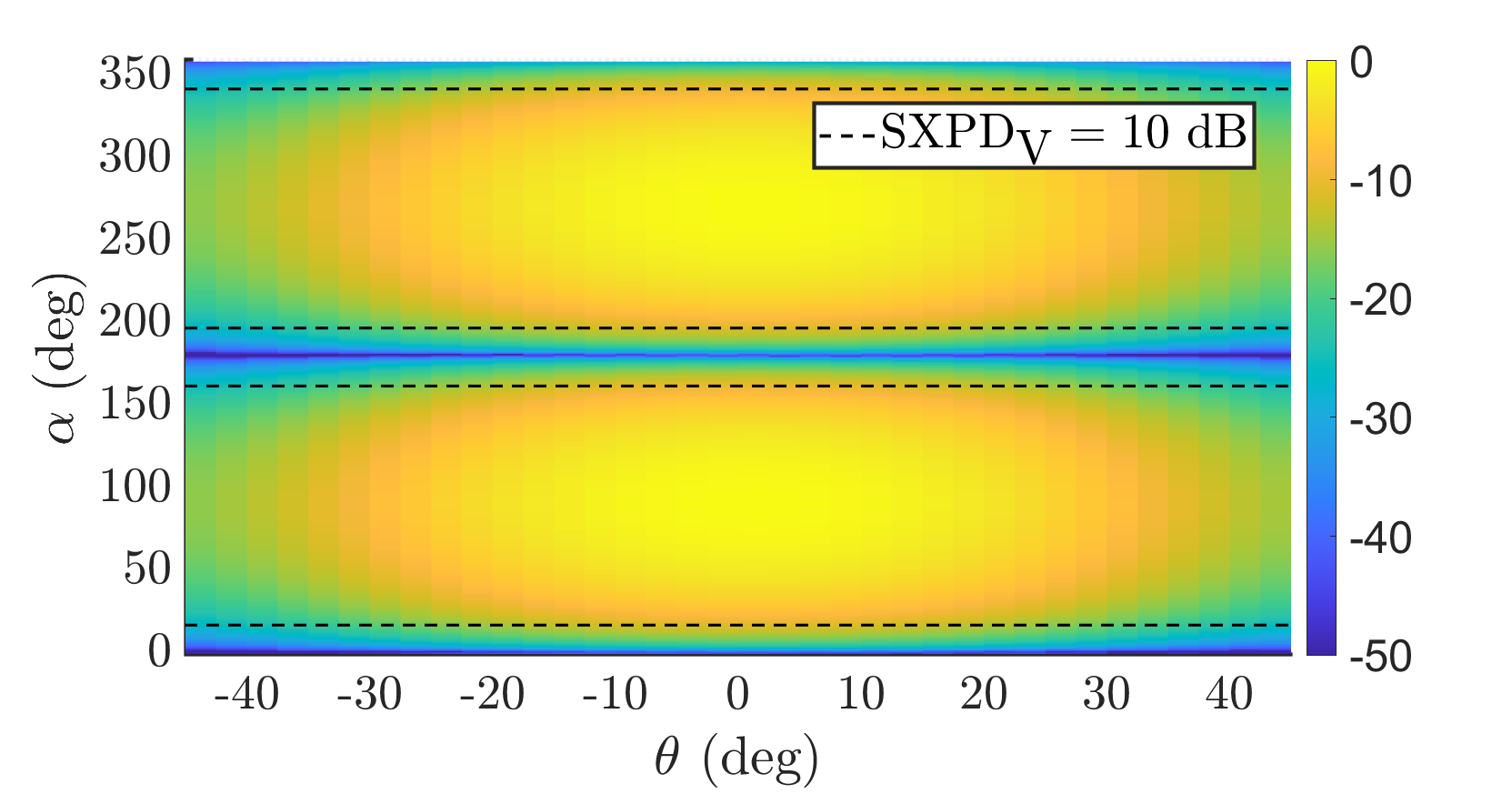}}
  \caption{Measured $\chi_{\textrm{VV}}$ and $\chi_{\textrm{VH}}$ for the dual-polarized antennas at $25$ GHz, marking $\alpha$ values for which $\textrm{SXPD}_{\textrm{V}} = 20 \log_{10} \left( \frac{\chi_{\textrm{VV}}}{\chi_{\textrm{VH}}} \right) = 10$ dB.}
  \label{chifig}
\end{figure*}

\subsection{Receiver}\label{rx_impl}

The receiver diagram is shown in Fig. \ref{txrxschemes}, with blue and red colors indicating digital signal processing and hardware blocks, respectively. Note that the architecture and decoding strategies are identical for the two receivers (users). After frequency-down conversion, the USRP performs low-noise amplification and ADC. Based on the alphabets $\mathcal{X}_0$ and $\mathcal{W}$, amplitude, time, frequency, and phase synchronization are performed using a digital automatic gain control (AGC), a polyphase-filter architecture \cite{poly} and a Costas control loop \cite{costas} combined with a numerically controlled oscillator (NCO). This process is explained in more detail in \cite[Section IV-C]{tcom}. Note that the approach used in this prototype is blind (i.e., no synchronization sequence is transmitted). In a final (commercial) implementation, the periodic transmission of a known sequence would reduce the number of packets lost before system convergence, which are not considered in our SER measurements. The decoders follow the expression presented in Section \ref{NSIC}. After symbol decoding, the indices corresponding to the private streams are selected before packet decoding. Each packet is detected via its access code, and its CRC32 header is evaluated for SER estimation.

\subsection{Final Deployment}
Fig. \ref{photos} presents photographs of the system components and the complete setup deployed in the anechoic chamber at the University of Oviedo, where the measurements reported in Section \ref{ER} were obtained. For simplicity, a single receiver is used to obtain the results for both users. Specifically, the transmitter is activated with a given configuration, and the receiver is configured to emulate user $ 1$. After this, measurements are taken. Subsequently, without modifying the transmitter configuration, the receiver is reconfigured to emulate user $2$, and the corresponding measurements are collected. This procedure is possible because both users share the same hardware configuration; the only difference lies in the software-based decoding process, as explained in Section \ref{rx_impl}. Note that, for the considered measurement campaign, in which each SER measurement is obtained under fixed channel conditions that do not vary temporally during the SER acquisition, the resulting measurements are expected to be identical to those obtained using two receivers simultaneously.

\section{Experimental Results}\label{ER}

\subsection{Polarization Measurements and General Considerations}\label{ERa}

Before deploying the communication system, we characterized the selected dual-polarized antennas in the anechoic chamber at the University of Oviedo. For each receive port (vertical and horizontal), we measure the captured signal components as functions of the azimuth and roll angles denoted by $\theta$ and $\alpha$ (see Fig. \ref{pos}), respectively. As a representative example, Fig. \ref{chifig} shows the normalized measured $\chi_{\textrm{VV}}$ and $\chi_{\textrm{VH}}$ at $25$ GHz, where dotted black lines indicate the values of $\alpha$ for which $\textrm{SXPD}_{\textrm{V}} = 20 \log{10} \left( \frac{\chi_{\textrm{VV}}}{\chi_{\textrm{VH}}} \right) = 10$ dB. Once all $\chi_{\textrm{PL}}$ components are measured, the desired $\textrm{SXPD}$ values can be accurately reproduced for the SER measurements presented in Section \ref{ERb}.

After the communications system is deployed, as shown in Fig. \ref{photos}, the total power given to both ports of the transmit antenna is approximately $-10$ dBm and is kept constant across all measurements. For example, when the power is equally split with $\gamma_0 = 0.5$, each port receives approximately $-13$ dBm. To enable a comparison between simulated and measured results, we define the SNR at the $i$-th receiver as $\textrm{SNR}_i = \displaystyle \frac{P_t |h_{i\textrm{PP}}|^2}{\sigma_i^2}$. In the measurement campaign, $\textrm{SNR}_i$ is estimated at each user by measuring the error vector magnitude (EVM) \cite[Eq. (16)]{tcom} when all transmit power is allocated to a single polarization using a $4$-QAM constellation and the same polarization is received. The procedure is then repeated for the other polarization to verify that the resulting $\textrm{SNR}_i$ is approximately equal for both polarizations. This behavior is expected due to the symmetric hardware configuration: identical cables and components for both polarizations in transmission and reception, and nearly identical radiation patterns for each polarization of each antenna. Fig. \ref{pos} shows the measurement campaign floor plan, depicting the different positions for the receiver, while the transmitter position is fixed. P1 and P2 receive approximately the same power level from the transmitter (and therefore exhibit similar EVM and estimated $\textrm{SNR}_i$ values), whereas position P3 is selected to demonstrate a 6 dB reduction in these parameters with respect to P1 or P2.

\subsection{SER Measurements}\label{ERb}

We now present the obtained SER measurements alongside the corresponding simulation results, considering three different cases noted as Case (i)-(iii). Simulated SER results are obtained through Monte Carlo simulations in Matlab considering an AWGN channel, based on the non-SIC decoding expression in Eq. (\ref{nosic_decod}). In all cases, we assume $\textrm{SXPD}_{i\textrm{V}} = \textrm{SXPD}_{i\textrm{H}} = \textrm{SXPD}_i$, which naturally corresponds to the proposed implementation, where the vertically and horizontally components always rotate jointly.

\begin{figure*}
\centering
 \subfloat[Case (i)\label{serc1}]{%
       \includegraphics[width=0.32\linewidth]{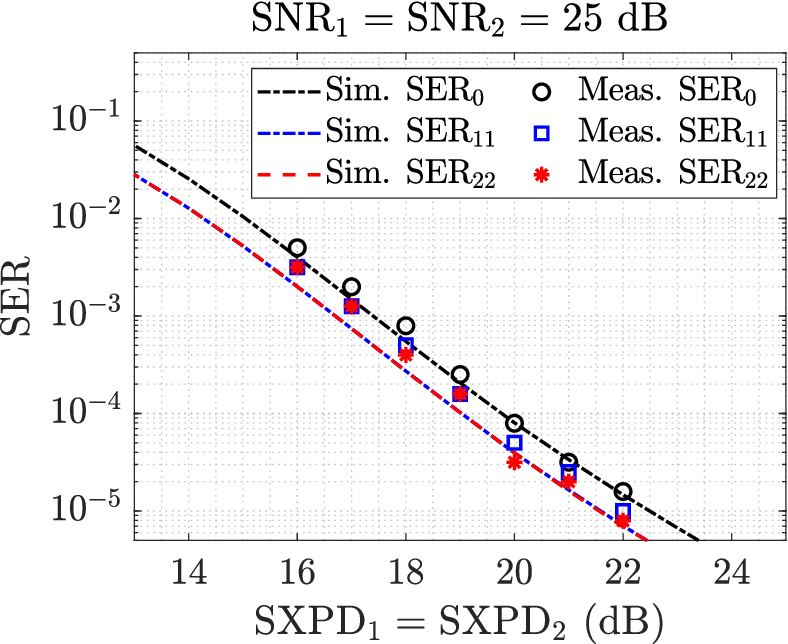}}
  \hfill
  \subfloat[Case (ii)\label{serc2}]{%
        \includegraphics[width=0.32\linewidth]{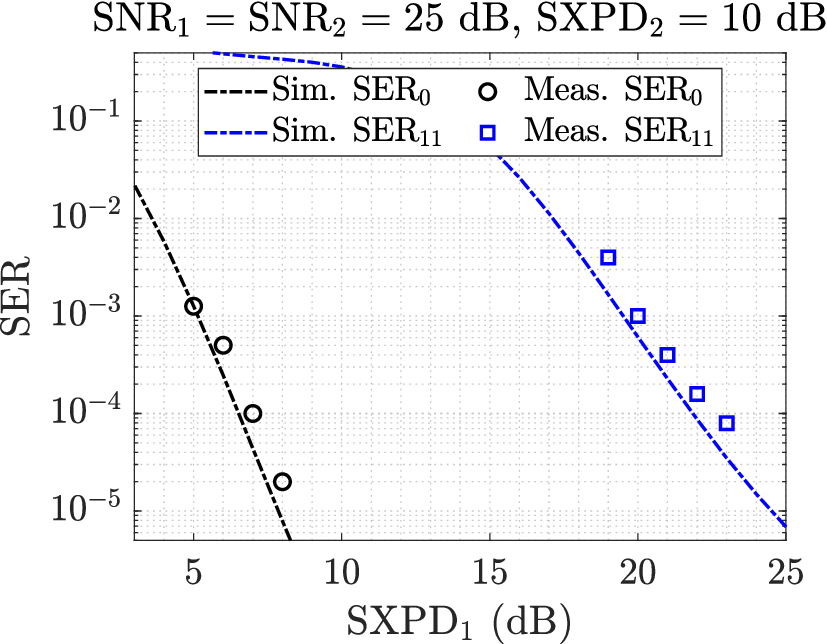}}
  \hfill
  \subfloat[Case (iii) \label{serc3}]{%
        \includegraphics[width=0.32\linewidth]{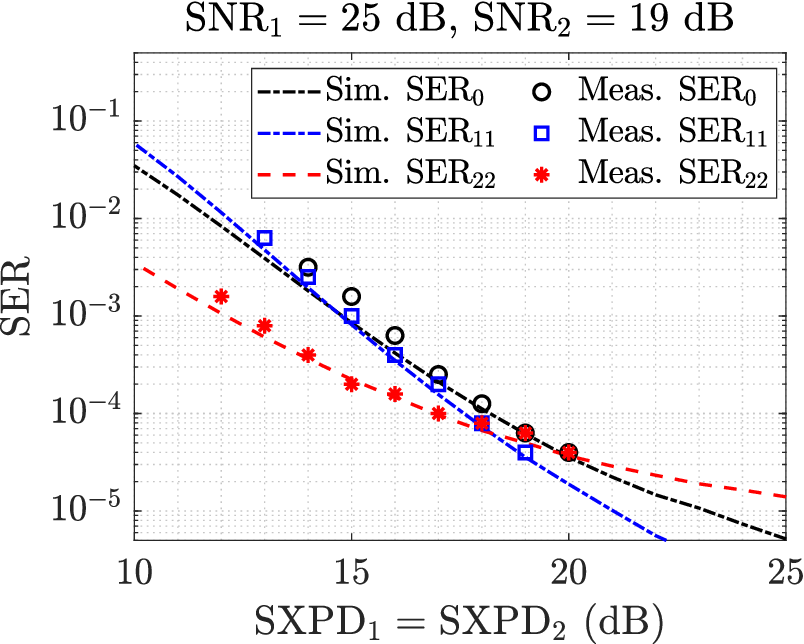}}
  \caption{SER measurements (Meas.) and simulations (Sim.).}
  \label{serres}
\end{figure*}

\begin{figure*}
\centering
 \subfloat[\label{Constellation_1}]{%
       \includegraphics[width=0.24\linewidth]{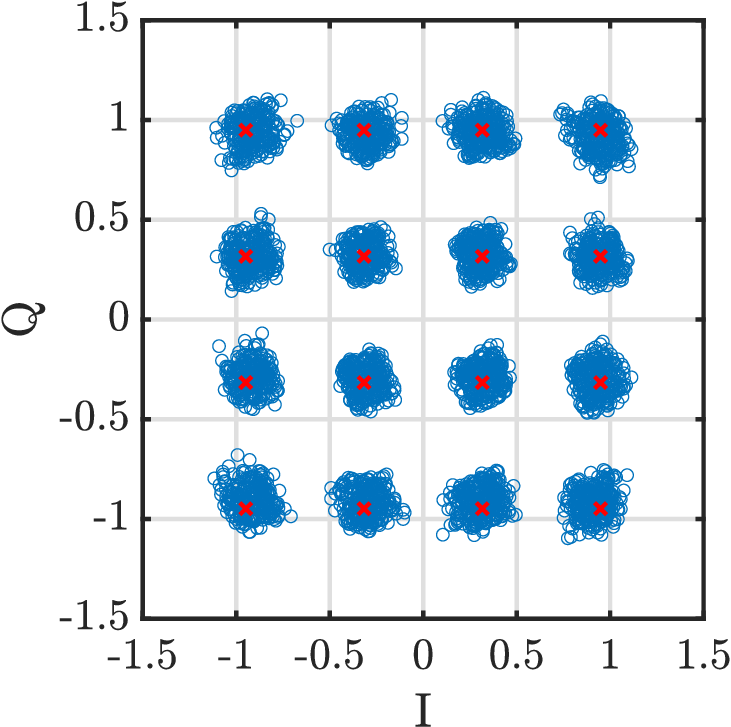}}
  \hfill
  \subfloat[\label{Constellation_2}]{%
        \includegraphics[width=0.24\linewidth]{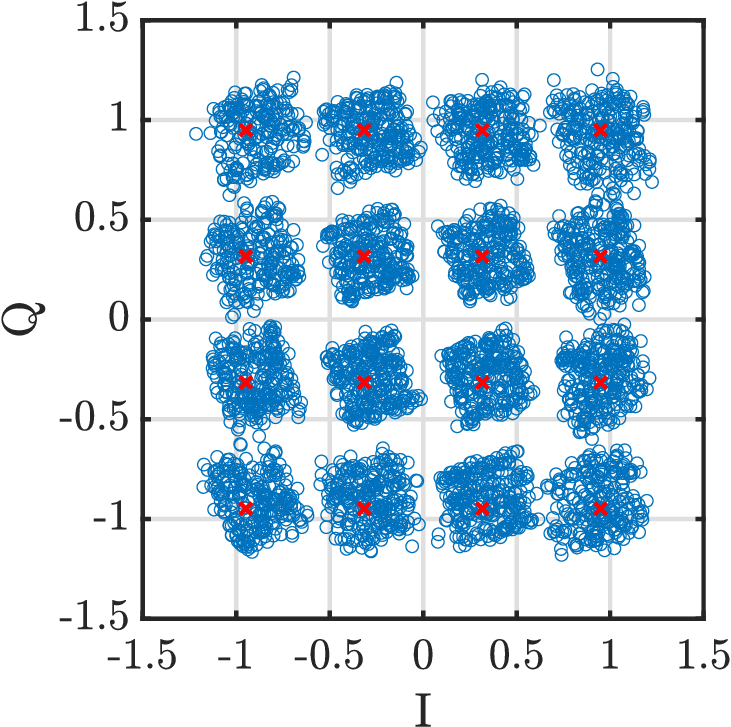}}
  \hfill
  \subfloat[\label{Constellation_3}]{%
        \includegraphics[width=0.24\linewidth]{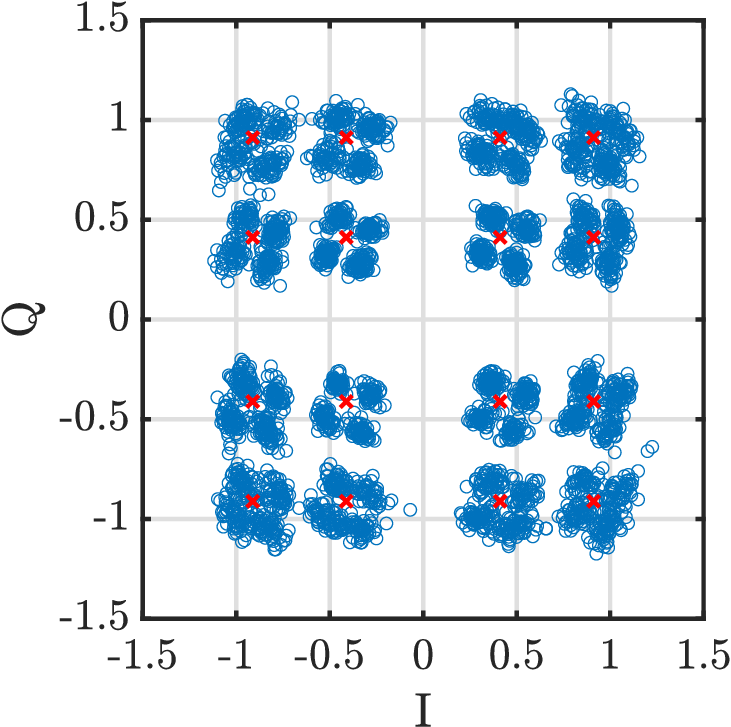}}
  \hfill
  \subfloat[\label{Constellation_4}]{%
        \includegraphics[width=0.24\linewidth]{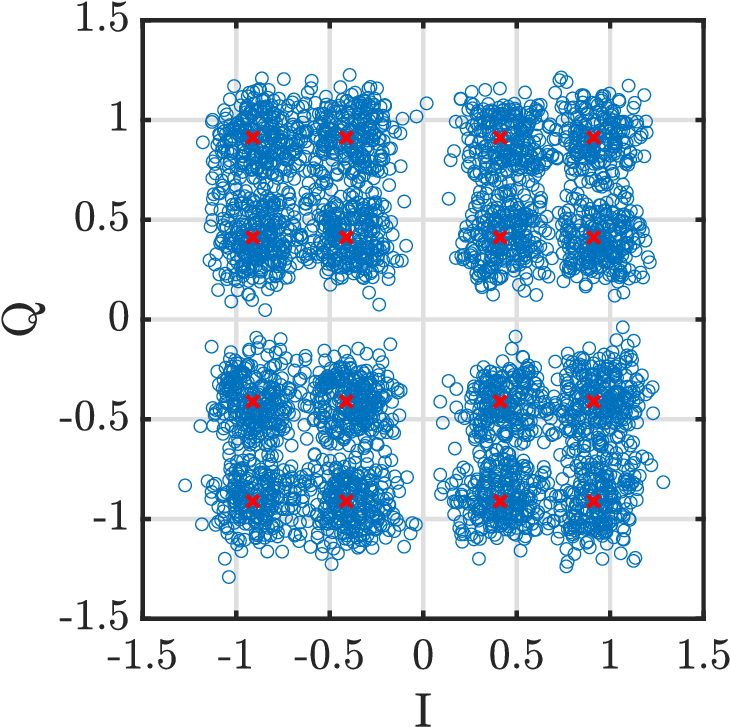}}
  \caption{Captured constellation symbols for: (a) Common stream constellation at user 1 (Case (i), position P1) for $\textrm{SXPD}_1 = 22$ dB; (b) Common stream constellation at user 1 (Case (i), position P1) for $\textrm{SXPD}_1 = 16$ dB; (c) Superposed constellation at user 1 (Case (iii), position P1) for $\textrm{SXPD}_1 = 15$ dB; (d) Superposed constellation at user 2 (Case (iii), position P3) for $\textrm{SXPD}_2 = 15$ dB. Transmitted symbols are marked with red crosses, and each figure displays 4000 received symbols. Measurement positions are indicated in Fig. \ref{pos}.}
  \label{constellations}
\end{figure*}

For Case (i), user $1$ and user $2$ are positioned at P1 and P2, respectively, so both users have similar channel conditions. Chosen constellations are $16$-QAM for the common stream, and $4$-PAM for the private streams, with one of them rotated by $90$ degrees, and a power allocation of $\gamma_0 = 0.5$, $\gamma_1=\gamma_2$. Note that this scenario aligns with the achievable rate results presented in Fig. \ref{fr2}. However, in this case, we fix the SNR for both users to $\textrm{SNR}_1 = \textrm{SNR}_2 = 25$ dB and present the SER results as a function of $\textrm{SXPD}_1 = \textrm{SXPD}_2$. During the measurements, $\textrm{SXPD}_i$ values can be obtained in a precise and reproducible manner thanks to the controlled setup in the anechoic chamber, as explained in Section \ref{ERa}. Therefore, particularly in the high-SNR regime, the good agreement observed in Fig. \ref{serc1} between measurements and simulations is expected. Remarkably, this RSMA configuration achieves a similar SER performance versus $\textrm{SXPD}_i$ for the three streams. Finally, Fig. \ref{Constellation_1} and Fig. \ref{Constellation_2} depict the captured constellation for the common stream (vertical polarization) at user $1$ (P1) for $\textrm{SXPD}_1 = 22$ dB and $\textrm{SXPD}_1= 16$ dB, respectively. In Fig. \ref{Constellation_2}, the interference caused by the horizontal polarization resulting from the superposed constellation of the private streams is clearly visible.

In Case (ii), user $1$ and user $2$ are also positioned at P1 and P2, respectively, with $\textrm{SNR}_1 = \textrm{SNR}_2 = 25$ dB. However, $\textrm{SXPD}_2$ is now fixed at $\textrm{SXPD}_2 = 10$ dB. Consequently, the private stream for user $2$ is turned off ($\gamma_2 = 0$). The common stream uses a $16$-QAM constellation with $\gamma_0 = 0.9$, while the private stream for user $1$ uses a $4$-QAM constellation with $\gamma_1 = 0.1$. Note that this scenario aligns with the achievable rate results presented in Fig. \ref{fr3}, where the objective is to provide additional data rate to user $1$ when its polarization conditions are favorable, while always protecting the common stream serving both users. This behavior is clearly observed in the results shown in Fig. \ref{serc2}.

Finally, in Case (iii) user $1$ is kept at P1, while user $2$ is positioned at P3, which results in different channel conditions, as considered previously in Section \ref{NR_DC}. Keeping the transmitted power constant, this scenario yields $\textrm{SNR}_1 = 25$ dB and $\textrm{SNR}_2 = 19$ dB, while we consider different values of $\textrm{SXPD}_1 = \textrm{SXPD}_2$. All streams use $4$-QAM constellations, with $\gamma_0 = 0.3$ and $\gamma_2 = 0.6125$, which sets $\gamma_1 = 0.0875$. The power allocation is selected by searching over the combinations $\gamma_0 \in \{0.1, 0.2, \ldots, 0.5\}$ and $\gamma_2 /(\gamma_2+\gamma_1) \in \{0.8, 0.825, \ldots, 0.9\}$ and choosing the pair that minimizes the maximum SER among the three streams at $\textrm{SXPD}_1 = \textrm{SXPD}_2 = 20$ dB. Note that the range $\gamma_2 /(\gamma_2+\gamma_1) \in [0.8, 0.9]$ has been previously shown to be of particular interest for the superposition of two $4$-QAM constellations \cite{ojcom}. A good agreement between measurements and simulations is again observed in the results presented in Fig. \ref{serc3}. A slight deviation appears at low $\textrm{SXPD}$ values, where the resulting interference begins to affect the receiver’s digital signal processing blocks, which are not modeled in the simulations, as perfect channel estimation and ideal time and frequency synchronization are assumed. Fig. \ref{Constellation_3} and Fig. \ref{Constellation_4} show the superposed constellation of both private streams for $\textrm{SXPD}_1 = \textrm{SXPD}_2 = 15$ dB at user $1$ and user $2$, respectively, where the SNR difference is clearly observed. Remarkably, user $2$ only needs to correctly detect the quadrant of the transmitted symbol to achieve correct decoding, which illustrates how the system can exploit channel disparities through appropriate power allocation.

\section{Conclusion}\label{Con}

In this work, we present the practical design and implementation of a dual-polarized RSMA system for mmWave communications. In particular, we consider a single transmitter and two receivers (users), each equipped with a pair of dual-polarized antennas. We first present the system and channel models and, inspired by previous theoretical works on GS, adopt a solution in which one polarization is used to transmit the common stream, while the superposition of the two private streams is transmitted over the other polarization. Remarkably, in the proposed design, the transmitter only requires knowledge of the channel strength and polarization conditions of each user (magnitude CSIT), and not of the instantaneous channel phase. Since our focus is on practical implementation, we  compute the achievable rate expressions for DS and evaluate them under different channel and polarization conditions for two receiver architectures: a SIC receiver, combining SIC and JMLD; and a non-SIC receiver, combining JMLD and TIN. Both architectures exhibit comparable performance for the specific case considered, provided that the transmitter is aware of the decoder architecture and adapts its transmission strategy accordingly.

We then present a prototype implementation of the system, based on SDR, using GNU Radio for digital signal processing and USRP devices (model X310) integrated with RF mixers and power amplifiers for the hardware components. The complete setup is deployed and tested in the anechoic chamber at the University of Oviedo. The dual-polarized antennas are first characterized so that the desired polarization conditions, specified by the SXPD ratios, can be accurately reproduced during the measurement campaign. SER measurements obtained under different channel conditions and SXPD values show good agreement with software simulations, thereby validating the system design and implementation. Although the experimental validation is conducted at $25$ GHz, the design can be readily adapted for operation in the upper portion of the envisioned FR3 band (e.g., between $20$ and $24$ GHz), where the considered channel model remains valid. The design may also be applicable to satellite communication downlinks because of both the channel characteristics and the emphasis on a limited number of transmit antennas.

Future research directions include integration with OFDMA and/or SC-FDMA \cite{rsma_ofdm}, further optimization of precoding and power allocation strategies \cite{rsma_finite3}, and the joint design of superposition and channel coding schemes to enhance BER performance \cite{Mod_Cod_NOMA}. 

To facilitate the reproducibility of this research, the code for the prototype implementation is available at https://github.com/alvpr/dualpolrsma.

\appendices
\section{Mutual Information Analysis} \label{appa}

The following analysis adopts a methodology similar to that used in \cite{rsma_finite} and \cite{superpo}, adapted here to the RSMA dual-polarized scenario without relying on the approximation based on Jensen's inequality used in \cite{rsma_finite}. Throughout all calculations, we assume infinite block length and that the symbols of each $k$-th stream, represented by the random variable $s_k$, are equiprobable. Under these assumptions, 
$H(s_k) = \log_2(M_k)$, where $H(s_k)$ denotes the entropy of $s_k$ \cite{book_in}. For the superposed constellation of the private streams 
$H(w)  = \log_2(M_1M_2) = \log_2(M_{12})$, where, as explained in Section \ref{txdes}, it is assumed that there are no repeated symbols in the superposed constellation. Finally, the notation $w_{ml}$, with $m \in \{1, 2, \ldots, M_1\}$ and $l \in \{1, 2, \ldots, M_2\}$ (Section \ref{txdes}), is used to denote a realization of $w$ explicitly indicating its corresponding source symbols, $s_{1m}$ and $s_{2l}$. In contrast, $w_q$ denotes a generic realization, where $q \in \{1, 2, \ldots, M_{12}\}$.  The power allocation values $P_k$, defined such that $x_k = \sqrt{P_k}s_k$, are treated as fixed and known constants. As a result, knowing the realization $s_{kn}$ directly determines $x_{kn}$, and vice versa.

\subsection{SIC with Perfect CSIR}
We first derive the expressions under the assumption of perfect SIC, which requires perfect CSIR. Note that in Section~\ref{SIC}, the symbol decoder is described using only the coefficients $h_{i\textrm{VV}}$, $h_{i\textrm{HH}}$, and $h_{i\textrm{HV}}$. Nonetheless, for the calculation of the constellation-constrained capacity, we assume that the $i$-th receiver knows all four channel coefficients, $\bold{H}_i$.

The $i$-th receiver first decodes the common stream in the vertical polarization at a rate upper bounded by the mutual information $R_{i0}^{\textrm{SIC}}=I(y_{i\textrm{V}};s_0|\bold{H}_i)$, and then, the $i$-th private stream after SIC, at a rate upper bounded by $R_{ii}^{\textrm{SIC}}=I(y_{i\textrm{H}};s_i|\bold{H}_i, x_0)$. We first focus on $R_{i0}^{\textrm{SIC}}=I(y_{i\textrm{V}};s_0|\bold{H}_i)$. Using the entropy and mutual information definitions \cite{book_in}:

\begin{equation}\label{pp2}
I(y_{i\textrm{V}};s_0|\bold{H}_i) = \operatorname{E}\left[ \log_2 \frac{p(y_{i\textrm{V}}|x_0,\bold{H}_i)}{p(y_{i\textrm{V}}|\bold{H}_i)}\right],
\end{equation} where $p(y_{i\textrm{V}}|\bold{H}_i)$, in a mild abuse of notation, denotes the probability density function (p.d.f) of $y_{i\textrm{V}}$ given $\bold{H}_i$, which can be expressed as 

\begin{align} \label{pp3}
&p(y_{i\textrm{V}}|\bold{H}_i) = \nonumber\\
&\frac{1}{M_0M_{12}} \sum_{n=1}^{M_0}\sum_{q=1}^{M_{12}}p(y_{i\textrm{V}}|x_0=x_{0n},w=w_{q},\bold{H}_i),
\end{align} where, for the sake of simplicity, the same terms $y_{i\textrm{V}}$ and $\bold{H}_i$ refer to the random variables and their realizations. For calculating the conditional p.d.f.  with AWGN, we have

\begin{align}\label{pp4}
&p(y_{i\textrm{V}}|x_0=x_{0n},w=w_q,\bold{H}_i) = \nonumber\\
&\frac{1}{\pi\sigma_i^2}\exp\left(-\frac{\left|y_{i\textrm{V}} - h_{i\textrm{VV}}x_{0n}-h_{i\textrm{VH}}w_{q}\right|^2}{\sigma_i^2} \right).
\end{align}

In the same manner, $p(y_{i\textrm{V}}|x_0=x_{0n},\bold{H}_i)$ is given by 

\begin{align} \label{pp5}
&p(y_{i\textrm{V}}|x_0=x_{0n},\bold{H}_i) = p(y_{i\textrm{V}}|x_{0n},\bold{H}_i) =\nonumber\\
&\frac{1}{M_{12}} \sum_{q=1}^{M_{12}}p(y_{i\textrm{V}}|x_0=x_{0n},w=w_{q},\bold{H}_i).
\end{align}

Combining Eq. (\ref{pp2})-(\ref{pp5}), and given that $ y_{i\textrm{V}} = h_{i\textrm{VV}}x_0 + h_{i\textrm{VH}}w + n_{i\textrm{V}}$, Eq. (\ref{I1}) is obtained in terms of $\operatorname{E}_{n_{i\textrm{V}},\bold{H}_i}[\cdot]$, which yields $R_{i0}^{\textrm{SIC}}$. Importantly, note that the channel and noise variables are independent. To calculate $R_{ii}^{\textrm{SIC}}$ after applying SIC, a similar methodology can be used, assuming knowledge of $x_0=x_{0n}$. The key difference is that the interfering stream is not $w$ as in Eq. (\ref{I1}), but the other private stream. Eq.(\ref{I2}) presents the expression for $R_{11}^{\textrm{SIC}}$, and $R_{22}^{\textrm{SIC}}$ can be straightforwardly obtained by adjusting the corresponding indices. 

Finally, note that expected value affects both the channel and noise coefficients, but knowing the distribution followed by $\bold{H}_i$, $n_{i\textrm{V}}$ and $n_{i\textrm{H}}$, numerical results can be readily obtained from these expressions using Monte Carlo simulations.

\subsection{Non-SIC with Partial CSIR}
We now consider the expression for a non-SIC receiver, based only on JMLD and TIN decoding and where each receiver chain only estimates $h_{i\textrm{VV}}$ and $h_{i\textrm{HH}}$ (Section \ref{NSIC}), which we denote as partial CSIR. Under this assumption, we model all interference as AWGN, defining the interference components, $z_{i\textrm{V}}\sim \mathcal{C}\mathcal{N}\left(0,\sigma_{z_{i\textrm{V}}}^2\right) $ and $z_{i\textrm{H}}\sim \mathcal{C}\mathcal{N}\left(0,\sigma_{z_{i\textrm{H}}}^2\right) $, where the variances are  given by

\begin{equation}\label{zvar}
\begin{cases}
\begin{aligned}
&\sigma_{z_{i\textrm{V}}}^2 = \sigma_i^2 + \operatorname{E}\left[|h_{i\textrm{VH}}|^2\right] (P_1+P_2)\\
&\sigma_{z_{i\textrm{H}}}^2 = \sigma_i^2 + \operatorname{E}\left[ |h_{i\textrm{HV}}|^2 \right] P_0.
\end{aligned}
\end{cases}
\end{equation}

Note that this modeling represents a worst-case assumption, as AWGN constitutes the worst type of interference in terms of mutual information, and it is unbounded, unlike the actual interference introduced by the information streams. For instance, in the horizontal polarization, the actual interference term is $h_{i\textrm{HV}}x_0$, where $x_0$ takes discrete values and $h_{i\textrm{HV}}$, although not known by the receiver in this case, is bounded. Under this modeling, similar steps to those used for the previous decoder are followed to compute $I(y_{i\textrm{V}}; s_0 | h_{i\textrm{VV}})$ and $I(y_{1\textrm{HH}}; s_1 | h_{1\textrm{HH}})$, resulting in $R_{i0}$ in Eq. (\ref{I3}), and $R_{11}$ in Eq. (\ref{I4}), respectively. Note that $R_{22}$ can be derived from Eq. (\ref{I4}) by appropriately adjusting the indices.

Due to the AWGN modeling, the resulting expressions are not exactly the mutual information between the input and output streams but rather conservative approximations. However, as already stated, this modeling is justified by the explicit choice of the decoder presented in Eq. (\ref{nosic_decod}). For instance, when decoding the private streams, note that the TIN strategy applied to the common stream and the lack of knowledge of the cross-polar channel coefficient, $h_{i\textrm{VH}}$, implicitly limit the possibility of exploiting the discrete structure of the interference. Thus, the presented approximations are more suitable for drawing conclusions about the potential of the proposed decoder.

\begin{strip}
  \rule{\textwidth}{0.4pt}
\begin{align}
    &R_{i0}^{\textrm{SIC}} = I(y_{i\textrm{V}};s_0|\bold{H}_i) = \log_2(M_0) \nonumber\\
    &-\frac{1}{M_0M_{12}}\sum_{n=1}^{M_0}\sum_{q=1}^{M_{12}}\operatorname{E}_{n_{i\textrm{V}},\bold{H}_i}\left[\log_{2}\left(\frac{\sum_{m=1}^{M_0}\sum_{l=1}^{M_{12}} \exp{\left(-\left| h_{i\textrm{VV}}(x_{0n} -x_{0m}) +   h_{i\textrm{VH}}(w_{q} - w_{l})+n_{i\textrm{V}}\right|^2/\sigma_i^2\right)}}{\sum_{l=1}^{M_{12}}  \exp{\left(-\left| h_{i\textrm{VH}}(w_{q} - w_{l})+n_{i\textrm{V}}\right|^2/\sigma_i^2\right)}}\right)\right] \label{I1}\\
    &R_{11}^{\textrm{SIC}} = I(y_{1\textrm{H}};s_1|s_0,\bold{H}_1) = \log_2(M_1) \nonumber\\
    &-\frac{1}{M_{12}}\sum_{n=1}^{M_1}\sum_{q=1}^{M_2}\operatorname{E}_{n_{1\textrm{H}},h_{1\textrm{HH}}}\left[\log_{2}\left(\frac{\sum_{m=1}^{M_1}\sum_{l=1}^{M_2} \exp{\left(-\left| h_{1\textrm{HH}}(x_{1n}  - x_{1m} + x_{2q} -x_{2l})+n_{1\textrm{H}}\right|^2/\sigma_1^2\right)}}{\sum_{l=1}^{l=M_2}  \exp{\left(-\left| h_{1\textrm{HH}}(x_{2q} - x_{2l})+n_{1\textrm{H}}\right|^2/\sigma_1^2\right)}}\right)\right] \label{I2}\\
    &R_{i0} = \log_2(M_0) -\frac{1}{M_0}\sum_{n=1}^{M_0}\operatorname{E}_{z_{i\textrm{V}},h_{i\textrm{VV}}}\left[\log_{2}\left(\frac{\sum_{m=1}^{M_0} \exp{\left(-\left| h_{i\textrm{VV}}(x_{0n} -x_{0m}) +z_{i\textrm{V}}\right|^2/\sigma_{z_{i\textrm{V}}}^2\right)}}{\exp{\left(-\left| z_{i\textrm{V}}\right|^2/\sigma_{z_{i\textrm{V}}}^2\right)}}\right)\right] \label{I3}\\
    &R_{11} = \log_2(M_1) \nonumber\\
    &-\frac{1}{M_{12}}\sum_{n=1}^{M_1}\sum_{q=1}^{M_2}\operatorname{E}_{n_1,h_{1\textrm{HH}}}\left[\log_{2}\left(\frac{\sum_{m=1}^{M_1}\sum_{l=1}^{M_2} \exp{\left(-\left| h_{1\textrm{HH}}(x_{1n}  - x_{1m} + x_{2q} -x_{2l})+z_{1\textrm{H}}\right|^2/\sigma_{z_{1\textrm{H}}}^2\right)}}{\sum_{l=1}^{l=M_2}  \exp{\left(-\left| h_{1\textrm{HH}}(x_{2q} - x_{2l})+z_{1\textrm{H}}\right|^2/\sigma_{z_{1\textrm{H}}}^2\right)}}\right)\right]  \label{I4}
\end{align}
  \vspace{10pt} 
  \rule{\textwidth}{0.4pt}
\end{strip}

\bibliographystyle{ieeetr}
\bibliography{references}

\end{document}